\documentclass[aps,prd,twocolumn,nofootinbib,showpacs,superscriptaddress]{revtex4-1}  % for review and submission
\usepackage{graphicx}  % needed for figures
\usepackage{dcolumn}   % needed for some tables
\usepackage{bm}        % for math
\usepackage{amssymb}   % for math
\usepackage{amsmath}
\usepackage{MnSymbol}
\usepackage[hidelinks]{hyperref}
\usepackage{graphicx}
\usepackage{subfig}
\usepackage{color}
\usepackage{dsfont}

\newcommand{\id}{\mathds{1}}

\begin{document}
\author{Mark Mace}
%\email{mark.mace@stonybrook.edu}
\affiliation{Physics and Astronomy Department, Stony Brook University, Stony Brook, NY, 11973, USA}
\affiliation{Physics Department, Brookhaven National Laboratory, Bldg. 510A, Upton, NY 11973, USA}
\author{S\"{o}ren Schlichting}
%\email{sschlichting@bnl.gov}
\affiliation{Physics Department, Brookhaven National Laboratory, Bldg. 510A, Upton, NY 11973, USA}
\author{Raju Venugopalan}
%\email{raju@bnl.gov}
\affiliation{Physics Department, Brookhaven National Laboratory, Bldg. 510A, Upton, NY 11973, USA}
\affiliation{Institut f\"{u}r Theoretische Physik, Universit\"{a}t Heidelberg, Philosophenweg 16, 69120 Heidelberg, Germany}
\title{Off-equilibrium sphaleron transitions in the Glasma}
\date{\today}
%\pacs{11.10.Wx,11.15.Ha, 12.38.Mh}
\begin{abstract} 
We perform first classical-statistical real time lattice simulations of topological transitions in the non-equilibrium Glasma of weakly coupled but highly occupied gauge fields created immediately after the collision of ultra-relativistic nuclei. 
Simplifying our description by employing SU(2) gauge fields, and neglecting their longitudinal expansion, we find that the rate of topological transitions is initially strongly enhanced relative to the thermal sphaleron transition rate and decays with time during the thermalization process. Qualitative features of the time dependence of this non-equilibrium transition rate can be understood when expressed in terms of the magnetic screening length, which we also extract non-perturbatively. A detailed investigation of auto-correlation functions of the Chern-Simons number ($N_{CS}$) reveals non-Markovian features of the evolution distinct from previous simulations of non-Abelian plasmas in thermal equilibrium. 
\end{abstract}
\maketitle

\section{Introduction}
Topological transitions are ubiquitous in nature and believed to be responsible for a variety of phenomena across the most diverse energy scales. In the context of non-Abelian gauge theories, a prominent example are transitions between energy degenerate ground states of definite Chern-Simons number, mediated by unstable, spatially localized classical field configurations of finite energy called Sphalerons~\cite{Klinkhamer:1984di,Manton:1983nd,Dashen:1974ck,Soni:1980ps,Boguta:1983xs}. Interest in these so-called sphaleron transitions first arose in the context of electroweak baryogenesis~\cite{Kuzmin:1985mm,Arnold:1987mh,Cohen:1993nk}: the high temperatures make these transitions energetically favorable and may lead to the large violations in baryon number necessary to explain the matter-antimatter asymmetry of the universe~\cite{Sakharov:1967dj}. 

Sphaleron transitions can also be significant in QCD at high temperatures and energy densities~\cite{McLerran:1990zh}. In this case, there is no baryon number violation;  however, due to the chiral anomaly and the Atiyah-Singer index theorem, these sphaleron transitions can generate significant amounts of axial charge. Because such transitions are governed by ultrasoft magnetic modes~\cite{Arnold:1987mh}, their rate in a non-Abelian plasma at finite temperature can be computed by solving classical equations of motion in real time~\cite{Grigoriev:1989ub}. Hard quantum modes still play a role due to Landau damping effects~\cite{Arnold:1996dy}; nevertheless, their effect can be accounted for in the hard thermal loop effective theory~\cite{Bodeker:1998hm} and the sphaleron transition rate can still be computed in a real time simulation non-perturbatively~\cite{Bodeker:1999gx}.  Estimates now exist for an SU(3) gauge theory at high temperatures where weak coupling methods are justified~\cite{Moore:2010jd}. 

A striking observation is that if the axial charge generated from sphaleron transitions is produced in the presence of a sufficiently strong external magnetic field, a net vector charge current can be produced in a hot QCD plasma~\cite{Kharzeev:2007jp,Fukushima:2008xe}. This phenomenon, called the Chiral Magnetic Effect (CME), can be studied in ultra-relativistic  heavy-ion collisions, where the magnetic fields in non-central collisions are very large at early times after the collision, and the energy densities are also sufficiently large that deconfined QCD matter is created. Experimental searches for the CME are ongoing at RHIC and the LHC, and intriguing hints suggestive of the CME and other anomalous transport effects have been seen \cite{Abelev:2009ac,Abelev:2012pa,Adamczyk:2014mzf,Adamczyk:2015eqo}. However conventional explanations for the observed signatures have also been put forward \cite{Schlichting:2010na,Schlichting:2010qia,Pratt:2010zn,Koch:2012kt,Hatta:2015hca} and further progress relies in part on an improved theoretical understanding of the expected magnitude and features of the signal. A status report on theoretical models of the CME and on-going experimental searches can be found in Ref.~\cite{Kharzeev:2015znc}. 

A major complication for theoretical descriptions of the chiral magnetic effect in heavy-ion collisions is the very short lifetime of the external magnetic field at the highest RHIC and LHC energies. Computations suggest that the magnitude of the magnetic field becomes smaller than the values relevant for the CME ($eB\geq m_\pi^2$) on very short time scales of $\sim 0.1$ to $0.2\, \frac{fm}{c}$ of the LHC and RHIC collisions respectively~\cite{Deng:2012pc} though in principle, very large electrical conductivities in the QGP can extend this time scale to slightly longer times~\cite{McLerran:2013hla}. Hence to understand whether the CME has observables consequences in heavy ion collisions, one needs to obtain an estimate of the sphaleron transition rate at very early times in the collision when the system is far off-equilibrium. 

A systematic QCD approach to the very early time evolution in ultra-relativistic heavy ion collisions is obtained in the Color Glass Condensate (CGC) effective field theory~\cite{Gelis:2010nm}. In this description, very high occupancy gluons in the nuclear wavefunctions are released in the collision generating non-equilibrium matter called the Glasma~\cite{McLerran:1993ni,Kovner:1995ts,Krasnitz:1998ns,Lappi:2006fp}. The dynamics of the Glasma is controlled by a saturation scale $Q_s$, which represents the hard momentum scale up to which gluons in the nuclear wavefunction have maximal occupancy. This saturation scale grows with the energy of the collision.  When $Q_s \gg \Lambda_{\rm QCD}$,  where $\Lambda_{\rm QCD}$ is the intrinsic QCD scale, the QCD coupling $\alpha_S(Q_s) <<1$ and weak coupling methods are applicable. Because the occupancy of gluons is parametrically of order $1/\alpha_S(Q_s) >> 1$ in the Glasma, the early time dynamics of this matter may be described by classical-statistical methods~\cite{Krasnitz:1999wc,Lappi:2003bi,Romatschke:2005pm,Epelbaum:2013waa,Berges:2013eia}.  One therefore has a clean theoretical limit in QCD, whereby the properties of the non-perturbative Glasma can be computed systematically~\cite{Gelis:2008rw,Lappi:2015jka}. 

We will present in this paper a first real time non-perturbative computation of the far from equilibrium sphaleron transition rate in the Glasma. For simplicity, we shall consider sphaleron transitions in a fixed box rather than the realistic (but computationally far more challenging) longitudinally expanding case~\footnote{Fluctuations in the Chern-Simons number for the longitudinally expanding, albeit boost invariant, Glasma were considered previously in \cite{Kharzeev:2001ev}. Because boost invariant gauge field configurations are 2+1-dimensional configurations, the second homotopy group of SU(2) is trivial. Thus in this case, non-zero integer valued topological transitions are not allowed; fractional values of topological charge can nevertheless be generated by fluctuations in the color electric and color magnetic fields.}. We will employ a numerical lattice implementation of the classical-statistical dynamics and adapt techniques previously developed in the context of classical Yang-Mills simulations to extract the thermal sphaleron transition rate~\cite{Ambjorn:1997jz,Moore:1996wn,Moore:1997cr,Moore:1998swa}, and in real-time studies of electroweak baryogenesis ~\cite{Smit:2002yg,Tranberg:2003gi,vanderMeulen:2005sp,Tranberg:2006ip,Tranberg:2006dg,Saffin:2011kn,D'Onofrio:2012jk}.

A deeper understanding of the rate of sphaleron transitions requires that one explore simultaneously the scales associated the hard modes as well as the softer electric and magnetic screening scales in the Glasma. How these scales develop with time has been discussed previously in the context of thermalization of the Glasma in both analytical~\cite{Blaizot:2011xf,Kurkela:2011ti} and numerical approaches~\cite{Berges:2007re,Schlichting:2012es,Kurkela:2012hp}. We will revisit this problem and demonstrate numerically that a clear separation of scales takes place with the temporal evolution of the Glasma. 

In particular, we will compute the spatial Wilson loop which provides the scale determining magnetic screening in a hot plasma in equilibrium. Albeit the non-equilibrium temporal evolution of the spatial Wilson loop has been studied previously~\cite{Berges:2007re}, we will go further and extract the scaling exponent that controls the temporal evolution of the corresponding string tension in the classical theory. We will show that the time evolution of the sphaleron transition rate is controlled by this string tension, scaling dimensionally as the string tension squared. Unlike the thermal case, topological transitions in the Glasma are determined by this magnetic scale alone and are robustly described by classical-statistical dynamics as long as occupancies are large.  

The outline of this paper is as follows: in Section \ref{overview}, we will give a brief overview of sphaleron transitions, in and out of equilibrium. In Section \ref{measuring}, we discuss in some detail how we measure the Chern-Simons number on a real time numerical lattice. We shall outline two different approaches, a ``slave field" approach and a ``calibrated cooling" approach and demonstrate, as a warm-up, our technology in the well understood case of thermal equilibrium initial conditions. Then in Section \ref{nonequilibrium}, we will introduce our non-equilibrium initial conditions and compute physical hard and soft scales as well as the far from equilibrium sphaleron transition rate. We will compare and contrast our results to those obtained in thermal equilibrium. In Section \ref{conclusion}, we will summarize our results, discuss their implications and outline future work. Some details of the numerical procedure and essential tests are described in two Appendices. 

\section{Sphaleron transitions and axial charge dynamics}
\label{overview}
In $SU(N_c)$ gauge theories with $N_f$ light flavors of fundamental fermions, the conservation of the axial current associated with each quark species 
\begin{eqnarray}
j^{\mu}_{5,f} = \overline{q_{f}} \gamma^{\mu} \gamma_{5} q_{f}\;,
\end{eqnarray}
is violated due to the axial anomaly as well as the explicit symmetry breaking introduced by the quark masses $m_f$
\begin{eqnarray}
\partial_{\mu}j^{\mu}_{5,f}= 2 m_{f} \overline{q} \gamma_{5} q - \frac{g^2}{16 \pi^2} F^a_{\mu \nu} \tilde{F}^{\mu \nu}_{a} \;,
\label{anomalyeqn}
\end{eqnarray} 
where $F^a_{\mu \nu}=\partial_\mu A^a_\nu-\partial_\nu A^a_\mu+g f_{abc}A^b_\mu A^c_\nu$ denotes the non-Abelian field strength tensor and $\tilde{F}^{\mu \nu}_{a} =\frac{1}{2}\epsilon^{\mu \nu \alpha \beta}F^a_{\alpha \beta} $ is its dual \cite{'tHooft:1976up}. Since in the high-temperature phase the explicit breaking due to the quark masses is usually neglected, we will denote the combined axial current of all light flavors as $j^{\mu}_{5}$ in the following. In this limit, the anomaly equation takes the form
\begin{eqnarray}
\label{anomaly-glue}
\partial_{\mu}j^{\mu}_{5}=-\frac{g^2 N_{f}}{16 \pi^2}  F^a_{\mu \nu} \tilde{F}^{\mu \nu}_{a} \;,
\end{eqnarray}
and states that fluctuations of the $SU(N_c)$ gauge fields characterized by $F_{\mu \nu} \tilde{F}^{\mu \nu} $ can induce local imbalances of the axial charge density $(j^{0}_{5})$ as well as global imbalances of the net axial charge
\begin{eqnarray}
J^{0}_{5}(t)=\int d^3 x~j^{0}_{5}(t,x)\;.
\end{eqnarray}
As noted previously, it was realized that in the presence of additional $U(1)$ electro-magnetic fields, such imbalances of axial charge densities $(j^{0}_{5})$ can lead to a variety of novel transport phenomena associated to the Chiral Magnetic (CME) \cite{Kharzeev:2007jp,Fukushima:2008xe} and related effects~\cite{Kharzeev:2015znc}.

Since the anomaly relation in Eq.~(\ref{anomaly-glue}) suggests that the fluctuations of axial charges are sourced by fluctuations of $F_{\mu \nu} \tilde{F}^{\mu \nu} $, basic features of the generation of an axial charge imbalance can be understood by investigating the non-Abelian dynamics of gauge fields\footnote{We note that in general there is a nontrivial interplay between the dynamics of the gauge sector and previously existing imbalances of axial charges \cite{Akamatsu:2013pjd,Akamatsu:2014yza,Akamatsu:2015kau}. However, since we are mostly interested in the generation of such axial charge imbalances we will not consider these effects within this study.}. Concentrating on the Yang-Mills sector from now on, it is convenient to express the right hand side of the anomaly equation in terms of the Chern-Simons current
\begin{equation}
\label{anomolycurrent}
K^\mu = \frac{g^2}{32 \pi^2}\epsilon^{\mu \nu \rho \sigma}\left(A^a_\nu F^a_{\rho\sigma}  - \frac{g}{3} f_{abc}A^a_\nu A^b_\rho A^c_\sigma \right)\;,
\end{equation} 
which satisfies the relation  
\begin{equation}
\partial_{\mu}K^{\mu}= \frac{g^2}{32 \pi^2} F^a_{\mu \nu} \tilde{F}^{\mu \nu}_{a} \; ,
\label{eq:CS-EB}
\end{equation}
such that the overall difference of net axial charge is given by
\begin{eqnarray}
\label{eq:NetChargeDifference}
J^{0}_{5}(t_2)-J^{0}_{5}(t_1)=-2 N_{f} \int_{t_1}^{t_2} dt \int d^3x~\partial_{\mu}K^{\mu}\;.
\end{eqnarray}
Since the spatial integral over the (four) divergence of the Chern-Simons current ($\int d^3x~\partial_{\mu}K^{\mu}$) defines a total time derivative, the right hand side of Eq.~(\ref{eq:NetChargeDifference}) can be expressed in terms of the difference of two boundary terms
\begin{eqnarray}
\int_{t_1}^{t_2} dt \int d^3x~\partial_{\mu}K^{\mu}=N_{CS}(t_2)-N_{CS}(t_1)\;,
\label{eq:CS-timediff}
\end{eqnarray}
where $N_{CS}$ denotes the Chern-Simons number
\begin{eqnarray}
N_{CS}(t)=\int d^3x~K^{0}(t,x)\;,
\label{ChernSimonsNumber}
\end{eqnarray}
which is a unique property of the gauge field configuration at the boundary. When considering a vacuum state for instance, the Chern-Simons number $N_{CS}$ can be associated with the homotopy class $Hom(G) \in \pi_{3}(SU(N_c)) \simeq\mathbb{Z}$ of the gauge transformation $G: \mathbb{R}^3 \cup \{\infty\} \to SU(N_c)$ that transforms the configuration to the topologically trivial vacuum state\footnote{While for gauge fields with periodic boundary conditions the gauge transformations map from the 3-torus $\mathbb{T}_{3}$ to the gauge group $SU(N_c)$, the set of homotopy classes of the map is still isomorphic to integers.}. (See for example \cite{Lenz:2001me} for a comprehensive review.) Consequently, the Chern-Simons number of a vacuum state is an integer which distinguishes between energy degenerate but topologically inequivalent configurations.

\begin{figure}
\centering
\includegraphics[width=0.2\textwidth]{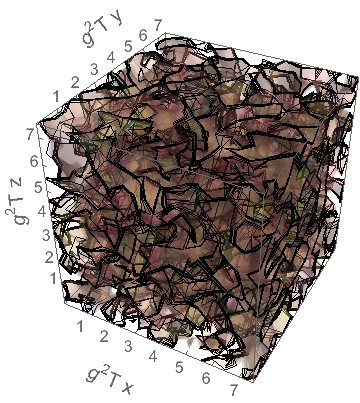}
\includegraphics[width=0.2\textwidth]{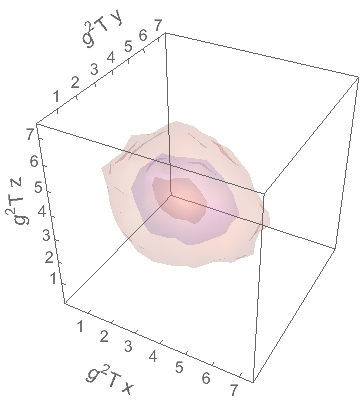}
\caption{\label{sphaleronmovie}Spatial profile of $\frac{g^2}{8\pi^2}~E^a_i(\mathbf{x} )B^a_i(\mathbf{x})$ for a 3D Yang-Mills configuration in thermal equilibrium ($N=16$,~$\beta=2$). Left: thermal field strength fluctuations contributing to $\vec{E}\cdot \vec{B}$ on all length scales. Right: spatial profile of $\vec{E}\cdot \vec{B}$ during a sphaleron transition after cooling to remove short distance fluctuations.}
\end{figure}

When considering states of finite energy density, either in or out of thermal equilibrium -- dynamical transitions between different topological sectors can occur~\cite{Kuzmin:1985mm}  mediated by the sphaleron~\cite{Klinkhamer:1984di}. While sphaleron transitions manifest themselves in a change of the gauge field topology, the Chern-Simons number of an excited state configuration is -- in contrast to vacuum states -- no longer necessarily an integer. Instead,  $N_{CS}(t)$ behaves as a continuous function of time whose derivative, according to Eqs.~(\ref{eq:CS-timediff})  and (\ref{eq:CS-EB}), satisfies the relation
\begin{equation}
\label{eq:NCsDeriviative}
\frac{dN_{CS}}{dt}=\frac{g^2}{8\pi^2}\int d^3x~E^a_i(\mathbf{x} )B^a_i(\mathbf{x})\;,
\end{equation}
where we will express $\frac{1}{4}F^a_{\mu \nu} \tilde{F}^{\mu \nu a} =E^a_i B^a_i$ with $E^{i}_{a}=F^{0i}_{a}$ and $B^{i}=\frac{1}{2} \epsilon^{ijk}F_{jk}$ from now on. Clearly, the integrand in Eq.~(\ref{eq:NCsDeriviative}) receives contributions not only from topological transitions but also from ordinary fluctuations of the field strength that are unrelated to topology. This is illustrated in Fig.~\ref{sphaleronmovie} where we show a snapshot of the integrand $\frac{g^2}{8\pi^2}   E^a_i(\mathbf{x} )B^a_i(\mathbf{x})$ for a thermal 3D Yang-Mills configuration. 
Thus from Eq.~(\ref{anomaly-glue}), both ordinary fluctuations and topological transitions contribute to the generation of an axial charge imbalance.

For Yang-Mills theories in thermal equilibrium, sphaleron transitions dominate the late time behavior of the Chern-Simons number \cite{Kuzmin:1985mm,Arnold:1987mh,Arnold:1987zg} that is related to the generation of a net axial charge. Since individual sphaleron transitions are uncorrelated with each other, the long time behavior of the Chern-Simons number in thermal equilibrium can be characterized by an integer random walk between different topological sectors. Accordingly, the sphaleron transition rate can be defined from the $N_{CS}$ auto-correlation function \cite{Khlebnikov:1988sr,Ambjorn:1995xm}
\begin{equation}
\Gamma_{sph}^{eq} \equiv \lim_{\delta t \rightarrow \infty} \frac{\langle(N_{CS}(t+\delta t)-N_{CS}(t))^2\rangle_{eq}}{V\delta t},
\label{rate}
\end{equation}
in the spirit of a transport coefficient. Since the definition of $\Gamma_{sph}^{eq}$ involves a real time correlation function, even in thermal equilibrium, this quantity is not accessible using first principles calculations in Euclidean lattice gauge theory. (See for instance \cite{Moore:2010jd} for a discussion.) Instead, calculations of the equilibrium sphaleron transition rate have been performed using either weak coupling numerical lattice techniques \cite{Grigoriev:1989ub,Ambjorn:1990wn} or by investigating theories with a holographic dual \cite{Son:2002sd}.

Parametric estimates of the sphaleron transition rate in  weakly coupled plasmas in thermal equilibrium give 
\begin{eqnarray}
\label{eq:ASYrate}
\Gamma_{sph}^{eq} = \kappa \;\alpha_S^5 T^4\;,
\end{eqnarray}
where $\alpha_S=g^2/4\pi$ here denotes the coupling constant \cite{Arnold:1996dy,Arnold:1998cy,Arnold:1999ux} and $\kappa$ is a non-perturbative constant. This parametric estimate is based on the argument~\cite{Kuzmin:1985mm,Arnold:1987mh,Arnold:1987zg} that the typical spatial length scale corresponding to a sphaleron transitions is determined by modes with wavelengths on the order of the inverse magnetic screening scale $1/g^2 T$.  The time scale necessary to achieve a sphaleron transition turns out not to be $1/g^2 T$ as previously argued~\cite{Arnold:1987zg}, but a longer time scale $1/g^4 T$, which accounts for the ``Landau damping" of the transition time due to the interaction of the soft magnetic modes with hard modes on the scale of the temperature. This parametric scaling has also been confirmed independently by numerical simulations \cite{Moore:1997sn,Moore:1999fs,Bodeker:1999zt,Moore:2010jd} which extract~\cite{Moore:2010jd} $\kappa= 0.21\, N_c^3 (N_c^2-1) (N_c g^2 T^2 /m_D^2)$ within logarithmic accuracy in the coupling $g$; plugging in the value of the Debye mass $m_D$ for very weak couplings, $N_f=0$ and $N_c=3$, one obtains $\kappa \sim 132\pm 4$. 

Our work here extends these studies to the case of an overoccupied non-Abelian plasma that is far off-equilibrium. Since such a system is dominated by classical dynamics, the sophisticated real time lattice techniques~\cite{Ambjorn:1997jz,Moore:1996wn,Moore:1997cr,Moore:1998swa} previously used to study sphaleron transitions in equilibrium can be straightforwardly adapted to the problem of interest here. However the problem in our case is simpler because, unlike the thermal case, hard quantum modes do not influence the sphaleron transition rate in the Glasma. The dynamics of non-equilibrium sphaleron transitions is entirely determined by the classical-statistical simulations; indeed, for the time scales studied, the sphaleron rate in the Glasma is governed by soft modes on the order of the magnetic screening scale determined from the spatial Wilson loop. Before we present these results, we will first review the real time numerical techniques that are essential to compute sphaleron transitions. 

\section{Topology measurement on the real time lattice}
\label{measuring}
In this section, we will outline the numerical classical-statistical methods that have been developed for the real time simulations of topological transitions 
at high temperatures and adapt these for our non-equilibrium context. One first solves the classical Yang-Mills equations on the lattice; in this first study, for simplicity, we will consider only the case of an SU(2) gauge theory. While a lattice formulation of the problem is essential to describe non-perturbative real-time phenomena, the lattice discretization, as we shall discuss, poses problems for the extraction of topological information in the plasma. We will outline the sophisticated methods that have been devised to reliably extract  Chern-Simons number on the lattice for non-Abelian plasmas in equilibrium. In particular, we will discuss the independent calibrated cooling and slave field methods and adapt these  to the non-equilibrium Glasma case. As a benchmark for our computations, we will reproduce and discuss key features of the well known equilibrium results. 

\subsection{Classical-statistical lattice setup}
\label{seq:latticesetup}
We discretize the theory on a 3D spatial lattice with $N$ sites and spacing $a$ in each direction following the Hamiltonian formulation of lattice gauge theory in temporal axial gauge. We define the lattice gauge link variables $U_{\mu}(x)$ and electric field variables $E^{\mu}(x)$ such that they transform according to
\begin{eqnarray}
E^{\mu}_{(G)}(x)&=&G(x)E^{\mu}(x)G^{\dagger}(x)\;, \nonumber \\ \qquad U^{(G)}_{\mu}(x)&=&G(x)U_{\mu}(x)G^{\dagger}(x+\hat{\mu})\;, 
\end{eqnarray}
under time independent gauge transformations. Defining the variation of the lattice gauge links with respect to the gauge fields as
\begin{eqnarray}
\frac{\delta U_{\mu}(x)}{\delta A_{\nu}^{a}(y)} = -iga~\tau^{a}~U_{\mu}(x) ~\delta_{\mu}^{~\nu} \frac{\delta_{x,y}}{a^3}\;,
\end{eqnarray}
where $\tau^a$ denotes the fundamental generator of $SU(N_c)$, we solve the classical Hamilton equations of motion
\begin{eqnarray}
\partial_{t} E^{\mu}_{a}(x) &=& -\frac{\delta H}{\delta A_{\mu}^{a}(x)}\;, \nonumber \\
 \partial_{t} U_{\mu}(x) &=& -iga~\tau^{a}~\frac{\delta H}{\delta E^{\mu}_{a}(x)}  U_{\mu}(x)\;,
 \label{eq:ClassicalHamilton}
\end{eqnarray}
derived from the lattice Hamiltonian
\begin{equation}
H=\frac{a^3}{2}\sum_{j,x} E^a_{j}(x) E^a_{j}(x) + \frac{2}{g^2 a}\sum_{\square}\mathrm{ReTr}\Big[\id-U_{\square}\Big]\;,
\end{equation}
using a leap-frog updating scheme. With the classical solutions in hand,  we can subsequently extract the observables of interest from the lattice field configurations and perform an average over an ensemble of all field configurations.

\subsection{Chern-Simons number measurement}
\label{sec:latticechernsimons}
While we are interested in the dynamics of topological transitions and the behavior of the Chern-Simons number over the course of the non-equilibrium evolution, there are several problems associated with the lattice definition of the corresponding observables--a clear discussion of these can be found in \cite{Moore:1998swa}. First of all, since the lattice is a discrete set of points, meaningful topological concepts can only be defined for i) sufficiently smooth configurations, for which gauge links are close to the identity and ii) slowly varying configurations, for which neighboring plaquettes are nearly identical. These are the gauge field configurations which effectively admit an interpolation between lattice sites. Further, there exists no local operator definition of the Chern-Simons current $K^{0}(t,x)$ such that the spatial integral $\int d^3x~K^{0}(t,x)$ is a total time derivative for generic field configurations. Instead,  local operator definitions of the Chern-Simons current can only be made approximately equal to a total derivative for sufficiently smooth and slowly varying field configurations.

Since these problems are particularly severe for classical-statistical simulations in thermal equilibrium, different techniques known as the ``calibrated cooling" \cite{Ambjorn:1997jz,Moore:1998swa} and the ``slave field" \cite{Woit:1985jz,Moore:1997cr} method have been developed in this context to overcome this challenge on real time lattices. We will briefly summarize below the basic ideas behind both methods. More details on the technical implementation of both methods are given in the Appendices~\ref{app::cooling} and \ref{app::sf}.

\subsubsection{Calibrated cooling}
\label{calibratedcoolingsection}
Problems with the lattice definition of the Chern-Simons current arise primarily due to ultraviolet fluctuations on the scale of the lattice spacing. However, fortuitously, the contribution of these ultraviolet modes to the sphaleron transition rate is suppressed because the sphaleron rate is dominated by the dynamics of modes on the order of the magnetic screening length \cite{Kuzmin:1985mm,Arnold:1987mh,Arnold:1987zg}. Hence an efficient way to deal with the aforementioned problems is to suppress the effect of ultraviolet fluctuations on the topology measurement by use of a ``calibrated cooling" technique \cite{Ambjorn:1997jz,Moore:1998swa}.

The most efficient way to remove ultraviolet fluctuations in a gauge invariant fashion is to follow the trajectory of the configuration along an additional ``cooling time" direction $\tau$ along which the gauge links evolve according to the energy gradient flow\footnote{While gradient flow techniques are also frequently used in Euclidean lattice gauge theory \cite{GarciaPerez:1994kz,deForcrand:1997esx,GarciaPerez:1998ru,Durr:2006ky,Luscher:2010we}, we note that the calibrated cooling technique employed here differs in the following aspects. While in 4D Euclidean lattice gauge theory, the gradient flow follows the steepest descent of the 4D gauge action, the cooling trajectory of our gauge field configuration follows the steepest descent of the 3D lattice Hamiltonian, such that three dimensional gauge field configurations at different times are cooled independently of each other.}
\begin{eqnarray}
\label{eq:UpdateStep}
\partial_{\tau} U_{\mu}(t,x;\tau) &=& -iga~\tau^{a} E_{cool}^{\mu,a}(t,x;\tau) \; U_{\mu}(t,x;\tau)\;, \\
E^{cool}_{\mu,a}(t,x;\tau) &=& - \frac{\delta H}{\delta A^{\mu}_{a}(x)}   \;.  \nonumber
\end{eqnarray}
By this procedure, ultraviolet fluctuations are efficiently removed and one can then define the Chern-Simons number by following the gradient flow (gf) all the way to the vacuum\footnote{Cooling to the classical vacuum amounts to following the energy gradient flow in Eq.~(\ref{eq:UpdateStep}) up to the point where the magnetic energy density vanishes identically to machine precision.}
\begin{eqnarray}
\label{eq:NCsVacCoolDef}
&& N_{CS}^{gf}(t)- N_{CS}^{vac}(t) =   \\
&& \qquad \qquad - \frac{g^2 a^3}{8 \pi^2} \int_{0}^{\infty} d\tau \sum_{x} E^{\rm cool}_{i,a}(t,x;\tau) ~ B^{\rm cool}_{i,a}(t,x;\tau)\;. \nonumber
\end{eqnarray}
While the Chern-Simons number of the associated vacuum configuration $N_{CS}^{vac}(t)$ is an integer characterizing the topological properties of the gauge field configuration, the integral on the right hand side contains the fluctuations above the vacuum and can be evaluated using an $\mathcal{O}(a^2)$ improved operator definition of the chromo-electric and chromo-magnetic fields inside the integral as described in Appendix~\ref{app::cooling}. Typically the field configurations become smooth already after cooling to $\tau a^2 \sim 1$. Thus ultraviolet lattice effects on the Chern-Simons number expression from integrating Eq.~(\ref{eq:NCsDeriviative}) are small and the method is topological \cite{Moore:1998swa}.  However the numerical cost of cooling all the way to the vacuum is too large for this definition to be practical and it is therefore useful to consider the following modifications instead.

\begin{figure}[t!]
\centering
\includegraphics[width=0.5\textwidth]{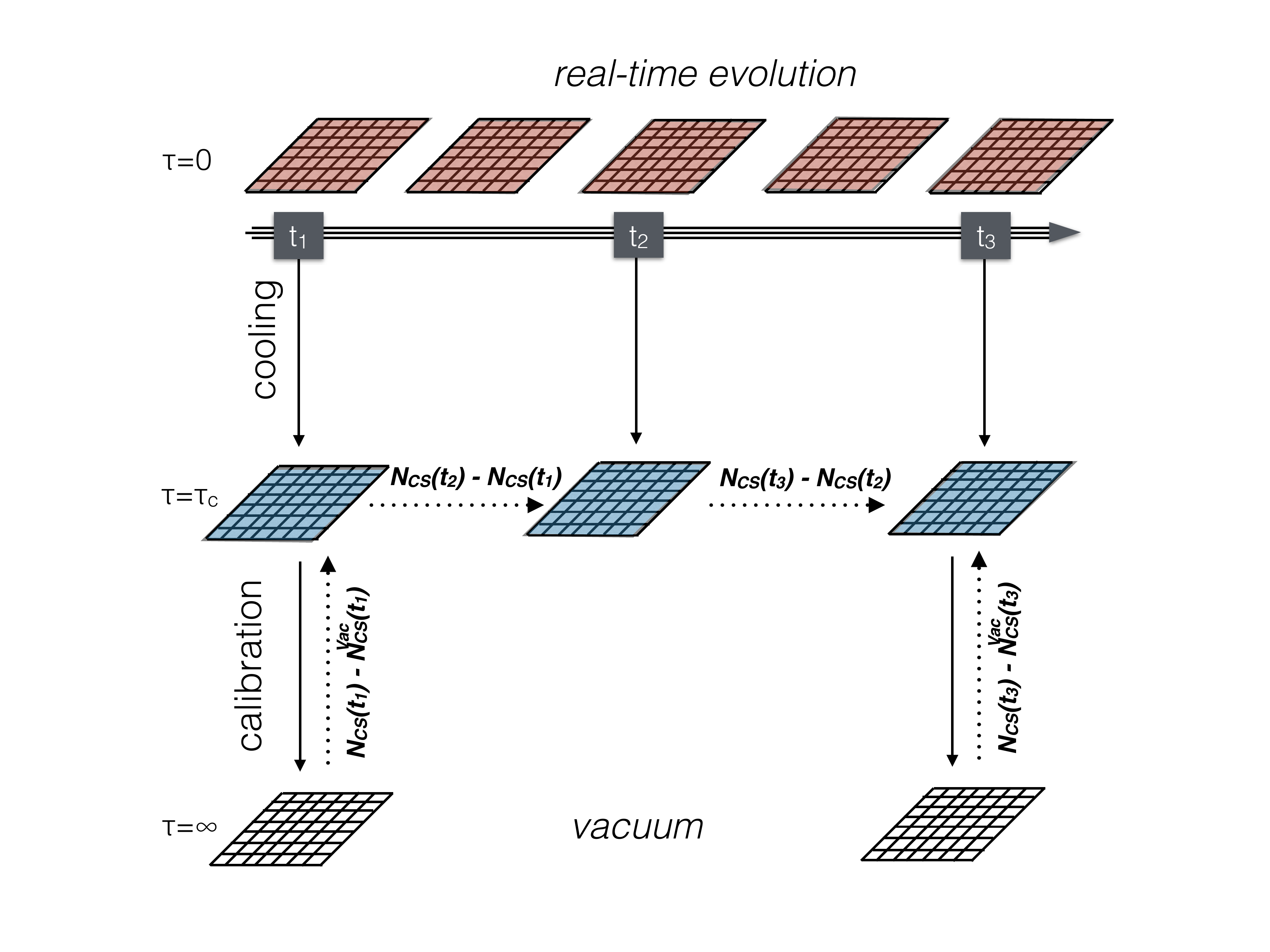}
\caption{Illustration of the calibrated cooling method.}
\label{fig:CoolingCartoon}
\end{figure}

Instead of cooling all the way to the vacuum $(\tau\to \infty)$ it is sufficient to cool for a shorter depth $\tau_c$ to efficiently remove ultraviolet fluctuations on the scale of a single lattice spacing. For such cooled configurations, a local operator definition of the Chern-Simons current behaves approximately as a total derivative; one can therefore define the change in Chern-Simons number between two different times $t_1$ and $t_2$ by comparing the cooled images of the configurations, 
\begin{eqnarray}
\label{eq:DeltaNCsCalibCool}
&& N_{CS}(t_2)-N_{CS}(t_1) = \\
&& \qquad \qquad  \frac{g^2 a^3}{8 \pi^2} \int_{t_1}^{t_2} d{t}' \sum_{x} E_{i,a}(t',x;\tau_c) ~ B_{i,a}(t',x;\tau_c)\;, \nonumber
\end{eqnarray}
as described in detail in App.~\ref{app::cooling}. By varying the amount of cooling $\tau_c$ applied to the configuration, one in addition controls  both the magnitude and typical wavelength of ordinary field strength fluctuations that contribute to the Chern-Simons number measurement. 
We shall  investigate the dependence on the cooling depth $\tau_c$ in more detail below. We note that one can check whether the definition of the Chern-Simons current indeed behaves as a total derivative by adding occasional coolings all the way to the vacuum to compute $N_{CS}$ (according to the previous definition in Eq.~(\ref{eq:NCsVacCoolDef})). This measurement is re-calibrated to ensure that any residual lattice errors do not accumulate over time. An illustration of this calibrated cooling method is shown in Fig.~\ref{fig:CoolingCartoon}.

In practice, we evolve the field configurations along the real-time axis $t$ and perform cooling up to a variable length $\tau_c$ for every 0.5 lattice units in time. Based on the cooled images of the configurations, we then compute the change in Chern-Simons number according to Eq.~(\ref{eq:DeltaNCsCalibCool}), as described in detail in Appendix~\ref{app::cooling}. We re-calibrate this measurement by performing a cooling to the vacuum\footnote{Since in practice the cooling to the vacuum is extremely costly, we use up to two steps of blocking \cite{Moore:1998swa} for our largest field configurations during this process. For more information, see Appendix~\ref{app::cooling}.} 
for  every 5 lattice units in time. When comparing the computations using respectively the definitions in Eqns.~(\ref{eq:DeltaNCsCalibCool}) and (\ref{eq:NCsVacCoolDef}), we observed that the largest discrepancy was of the order of 0.1 (and typically much smaller than that). The close agreement indicates that the method employed is indeed topological.

\subsubsection{Slave Field Method}
\label{slavefieldsection}
In the calibrated cooling method, ultraviolet fluctuations are removed explicitly to arrive at a topological definition of the Chern-Simons number. An  alternative method referred to as the ``slave field" method was proposed by Woit \cite{Woit:1985jz} and developed further by Moore and Turok \cite{Moore:1997cr}. The basic idea underlying this method is to extract the integer part of the Chern-Simons number (which, loosely speaking, corresponds to the integer part $N_{CS}^{vac}(t)$ in Eq.~(\ref{eq:NCsVacCoolDef})) by measuring the winding number $N_{W}$ of the gauge transformation $S(x)$ which transforms the configuration to the topologically trivial sector, setting
\begin{eqnarray}
N_{CS}^{SF}=N_W\;,
\end{eqnarray}
where the acronym SF stands for Slave Field. Since the topological trivial sector of 3D Yang-Mills theory satisfies the (minimal) Coulomb gauge condition that 
\begin{eqnarray}
Q=\frac{1}{3 N^3 N_c}\sum_{j,x} \text{ReTr} \Big[\id - U_{j}^{(S)}(x)\big]
\end{eqnarray}
is minimal, finding the gauge transformation $S(x)$ is then equivalent to the problem of fixing Coulomb gauge on the real time lattice. As pointed out in \cite{Moore:1997cr}, keeping track of this gauge transformation over the course of the Hamiltonian time evolution can be efficiently achieved using a sequence of small gauge transformations determined by an extended variant of the Los Alamos gauge fixing algorithm described in more detail in Appendix \ref{app::sf}.

At first sight, it may appear as if this procedure were completely free of the aforementioned ultraviolet problems. However these problems return in  determining the winding number of the gauge transformation. Indeed, a topologically meaningful definition of the winding number is only possible when $S(x)$ is sufficiently slowly varying\footnote{The slave field $S(x)$ does not necessarily have to be close to the identity for this condition to be satisfied.}; when this is the case, the winding number can be extracted using the methodology of Woit \cite{Woit:1985jz}. Specifically, for the $SU(2)$ gauge group, the winding number is characterized by the degree of the map
\begin{eqnarray}
N_W=deg\;(S)
\end{eqnarray}
and can be extracted in a straightforward way following \cite{Moore:1997cr}--the procedure is described in further detail in Appendix~\ref{app::sf}. Similarly, in the case of $SU(N_c)$ with $N_c>2$ , one can extract the winding number by decomposition into $SU(2)$ sub-groups~\cite{Parisi:1984nx}.

By fixing the Coulomb gauge condition at the initial time, the slave field $S(x)$ can initially be set to the identity such that it is slowly varying at that time. However, ensuring that the slave field $S(x)$ remains slowly varying over the course of the Hamiltonian evolution is a non-trivial task and generally requires a more careful tuning of the algorithm. The strategy devised in \cite{Moore:1997cr} is to monitor the roughness of the gauge transformation. This quality is quantified in terms of the peak stress defined as 
\begin{eqnarray}
PS=\text{sup}_{x} \sum_{j} \text{ReTr}\Big[\id -  \frac{1}{2} \Big( U_{j}^{(S)}(x) + U_{j}^{\dagger,(S)}(x-\hat{j}) \Big) \Big] \;, \nonumber \\
\end{eqnarray}
and to apply the transformation to Coulomb gauge of the dynamical fields $U$,~$E$ and $S$
\begin{eqnarray}
U_{i}(x) \to U_{i}^{(S)}(x)\;, \quad  E_{i}(x) \to E_{i}^{(S)}(x)\;, \quad S(x) \to \id \;, \nonumber \\
\end{eqnarray}
whenever the peak stress falls below a certain threshold~\cite{Moore:1997cr}. Consequently, by restoring the Coulomb gauge condition dynamically, one ensures that the slave field remains slowly varying on large time scales and the extraction of the winding number is topological.

Despite the elegance of the slave field method, in practice, the underlying local gauge fixing algorithm is very inefficient on large lattices and performing the frequent gauge fixing required for the topology measurement becomes prohibitively expensive. We will therefore use this method primarily as a benchmark and cross-check of the calibrated cooling technique discussed previously.

\subsection{Chern-Simons measurements \& sphaleron transitions in thermal equilibrium}
\label{sec:thermal}
Before we apply the above methods to study sphaleron transitions in the Glasma, we will briefly discuss the application of these methods to $SU(2)$ Yang-Mills theory in thermal equilibrium. While the correct determination of the sphaleron rate at weak coupling requires a simultaneous description of the hard ($\sim T$) and soft ($\sim g^2 T$) excitations of the systems -- as discussed earlier in Sec.~\ref{overview} -- our primary goal is to illustrate basic features of the evolution of the Chern-Simons number in thermal equilibrium. We will therefore neglect the effects of hard excitations in the thermal case and instead follow previous work that explored the behavior of soft ($\sim g^2 T$) modes in 3D classical Yang-Mills theory.

We generate thermal configurations using the highly efficient thermalization algorithm developed by Moore \cite{Moore:1996qs}. Beginning with a cold start ($U_{\mu}=\id\;, E^{\mu}=0$), we perform a series of iterations of the following steps:
\begin{itemize}
\item[1)] Generate color electric fields according to a Gaussian distribution with
\begin{eqnarray}
\langle E^{i}_{a}(x) E^{j}_{b}(y) \rangle = \beta_{lat} ^{-1} \delta^{ij} \delta_{ab} \delta_{x,y} \nonumber
\end{eqnarray}
where $\beta_{lat}=1/(g^2 T a)$ is the lattice coupling in 3D Yang-Mills theory\footnote{Our normalization of the lattice coupling differs by a factor of $4$ from the one used in earlier works \cite{Moore:1997cr}.}.
\item[2)] Project the color electric fields on the constraint surface where Gauss's law condition $D_{\mu}E^{\mu}=0$ is satisfied using the algorithm described in \cite{Moore:1996wn}. 
\item[3)] Evolve the gauge links and electric fields according to Hamilton's equations of motion for some time $t_{th}$ to allow energy to be exchanged between the electric and magnetic fields.
\end{itemize}
until the energy density settles to its equilibrium value\footnote{For a choice of $t_{th} a=50$  we find that for a $N=24$ lattice and $\beta_{lat}$ values as discussed below, convergence is typically achieved in less than $20$ iterations.}.  We subsequently perform the real time evolution by  solving Hamilton's equations of motion as described in Sec.~\ref{seq:latticesetup} and measure the Chern-Simons number using the methods described in Sec.~\ref{sec:latticechernsimons}. Since our primary aim was to illustrate the time evolution of the Chern-Simons number in the equilibrium setup, we performed our thermal simulations on rather small $N=24$ lattices with $\beta_{lat}=2$ where comparisons to published results are available \cite{Moore:1997cr}. 

\begin{figure}[t!]
\centering
\includegraphics[width=0.5\textwidth,natwidth=610,natheight=642]{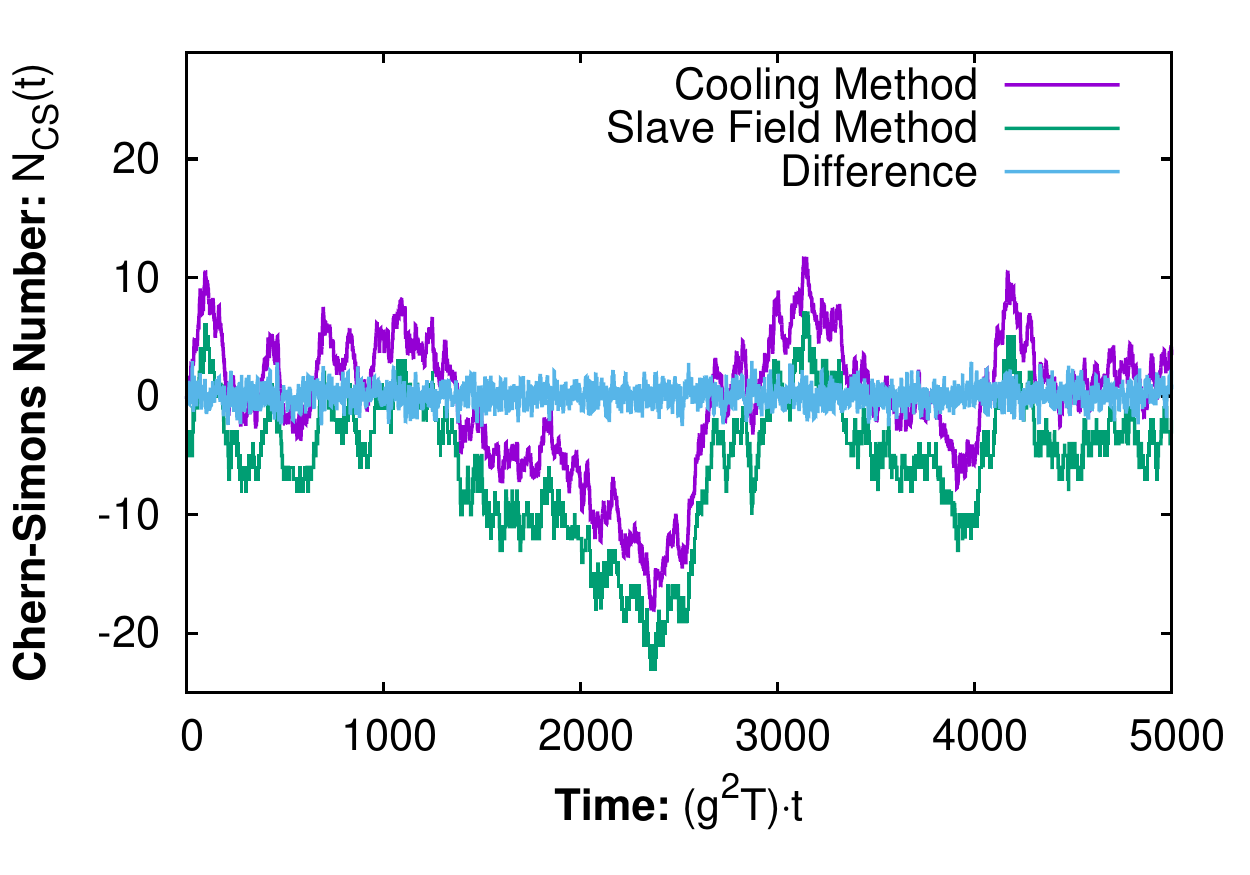}
\caption{Comparison of Chern-Simons number measurements using the calibrated cooling ($\tau_c=0.75~(g^2T)^{-2}$) and slave field techniques for a single configuration in thermal equilibrium ($N=24$, $\beta=2$).  Slave field results are shifted by 5 units for comparison purposes.  The difference between the two methods is small over the course of the entire simulation.}
\label{thermalsphaleron}
\end{figure}

We first investigate the behavior of the Chern-Simons number over the course of the real time evolution for a single field configuration. Our results are shown in Fig.~\ref{thermalsphaleron}. The different curves correspond to the extraction of the Chern-Simons number using the calibrated cooling (purple) and slave field (green) techniques. Since the difference between the two measurements -- also shown in Fig.~\ref{thermalsphaleron}  -- is small, we have shifted the slave field measurement by $5$ units for better comparison of the results.

The long time behavior of the correlation function in Fig.~\ref{thermalsphaleron} is dominated by transitions between different topological sectors that are characterized in terms of approximately integer changes of the Chern-Simons number over a short amount of time. While thermal field strength fluctuations also contribute to the calibrated cooling measurement, the comparison with the slave field measurement -- designed to measure  the topological contribution only -- shows that the effect of ordinary field strength fluctuations is small and does not significantly affect the long time behavior of the Chern-Simons number. We find, consistent with the results of \cite{Moore:1997cr,Moore:1998swa}, that both methods agree within statistical white noise over the course of the entire simulation.

\begin{figure}
\includegraphics[width=0.5\textwidth,natwidth=610,natheight=642]{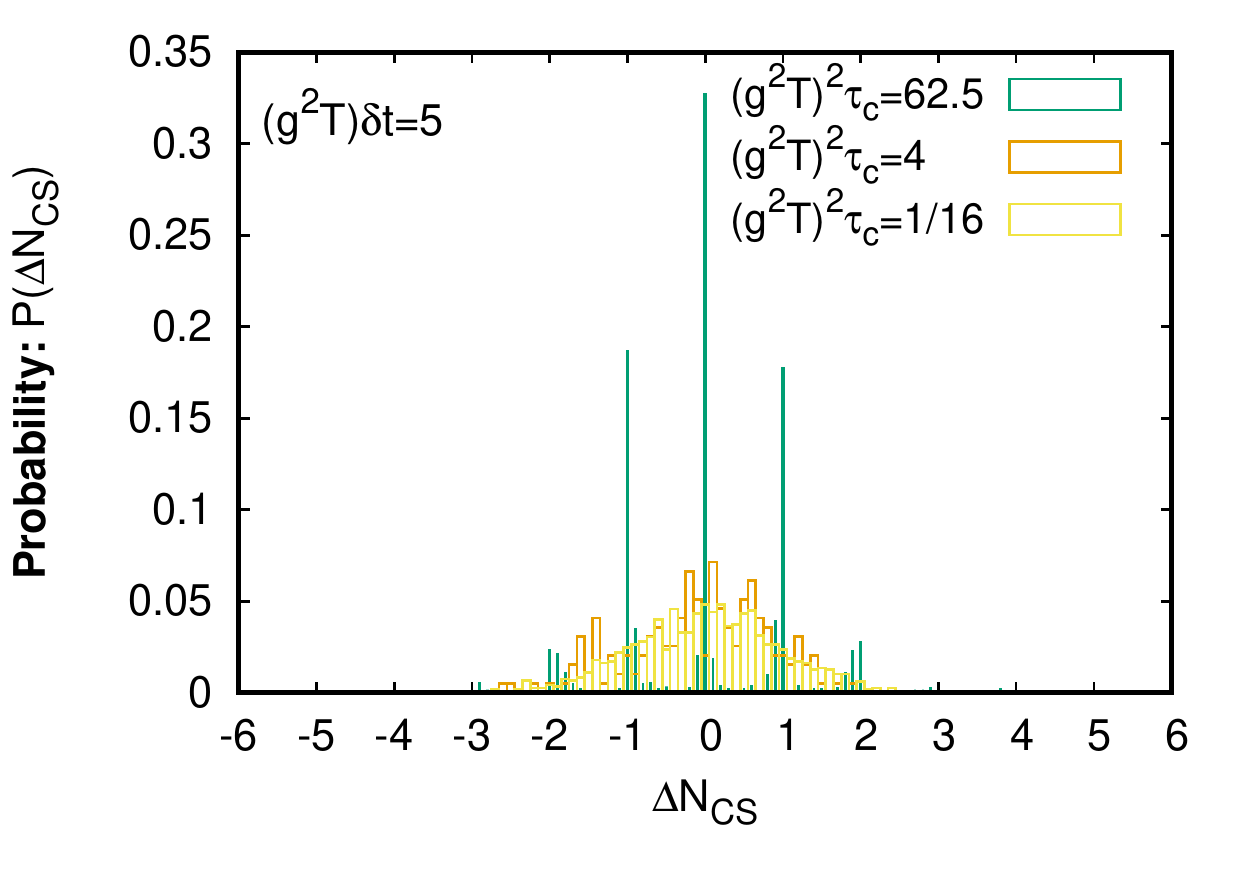}
\newline
\includegraphics[width=0.5\textwidth,natwidth=610,natheight=642]{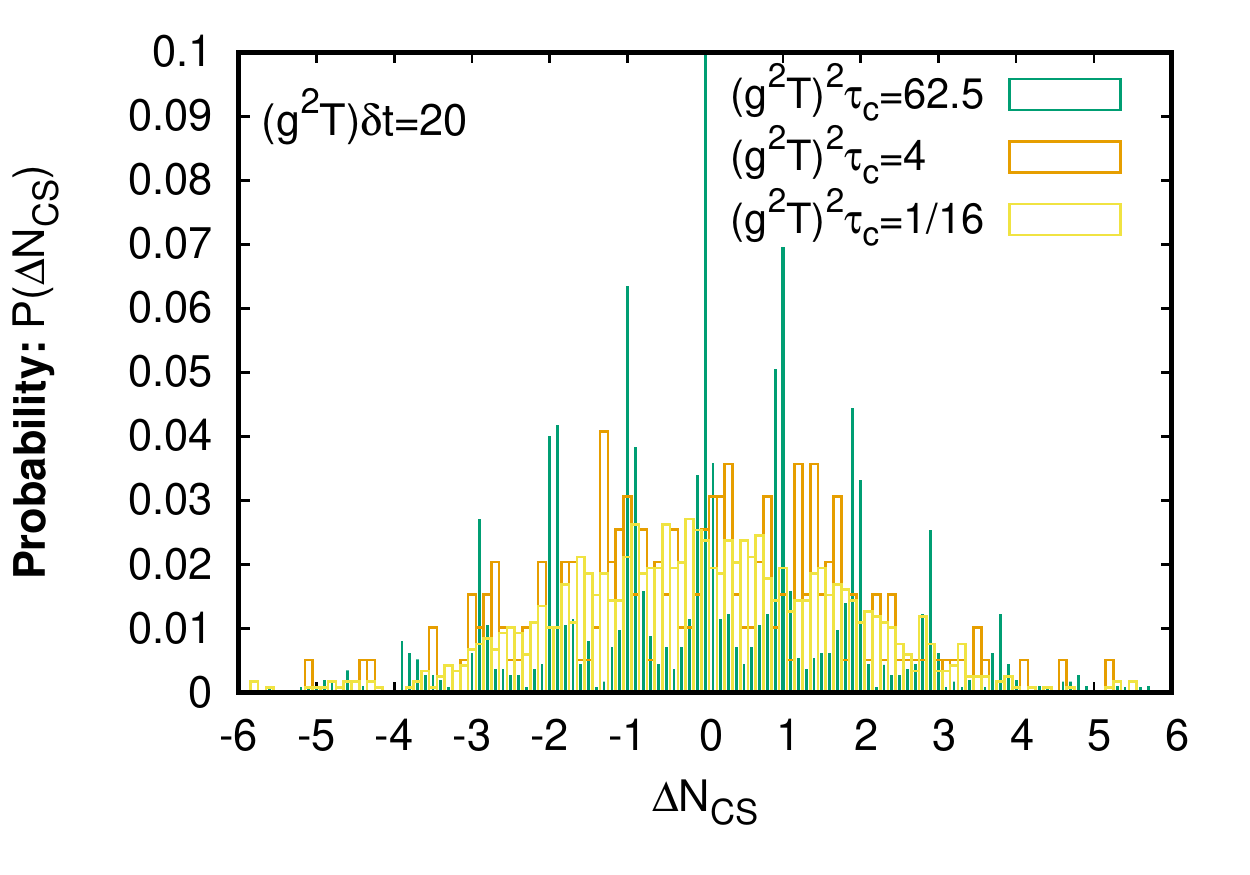}
\caption{Histogram of the Chern-Simons number difference $\Delta N_{CS}$ for equilibrium configurations  ($N=24$, $\beta=2$) separated by  $ \delta t = 5~(g^2 T)^{-1}$ (upper panel) and respectively $\delta t = 20~(g^2 T)^{-1}$ (lower panel) during the real time evolution. The different curves in each panel correspond to the results for different levels of cooling.}
\label{thermalhistograms}
\end{figure}

We can further quantify the time evolution of the Chern-Simons number by investigating the statistical properties of the Chern-Simons number difference
\begin{eqnarray}
\label{eq:DeltaNCSDef}
\Delta N_{CS}(t,\delta t)=N_{CS}(t+\delta t) - N_{CS}(t)\;,
\end{eqnarray} 
during the real time evolution. In equilibrium,  time translation invariance guarantees the result to be independent of $t$ and we have verified this explicitly. 
Our results for the probability distribution $P(\Delta N_{CS})$ are shown in Fig.~\ref{thermalhistograms} based on the data sets in Tab.~\ref{tab:EqData}. Different panels in Fig.~\ref{thermalhistograms}  show our results for two different separations in time $\delta t=5~(g^2T)^{-1}$ (upper panel) and $\delta t=20~(g^2T)^{-1}$ (lower panel), whereas different curves in each panel correspond to different choices of the cooling depth $\tau_{c}=0.0625 - 62.5 ~ (g^2T)^{-2}$ employed in the measurement of the Chern-Simons number.

While a minimal amount of cooling is necessary to ensure that the Chern-Simons current behaves at least approximately as a total time derivative\footnote{The quality of this approximation can be checked explicitly during the calibration step \cite{Moore:1997cr}. We refer to Appendix~\ref{app::cooling} for a more detailed discussion.}, by cooling further one successively removes residual field strength fluctuations from the configurations  and thereby reduces their contribution to the Chern-Simons number. After cooling down to $\tau_{c}=62.5 ~ (g^2T)^{-2}$, the difference in Chern-Simons number is strongly peaked around integers as the measurement is completely dominated by transitions between different topological sectors\footnote{Small differences from integer values arise not only due to residual field strength fluctuations, but also due to (small) lattice discretization errors in the determination of the space-time integral of the Chern-Simons current in Eq.~(\ref{eq:DeltaNCsCalibCool}).}. Since more transitions occur over a larger time scale $\delta t$, the width of the distribution increases significantly,  as is clear from the results shown in the upper and lower panels of Fig.~\ref{thermalhistograms}. Even though this behavior is most prominent for the coolest configurations, it can also be clearly seen for modest cooling times relative to the behavior of the original field configurations.

\begin{figure}
\centering
\includegraphics[width=0.5\textwidth,natwidth=610,natheight=642]{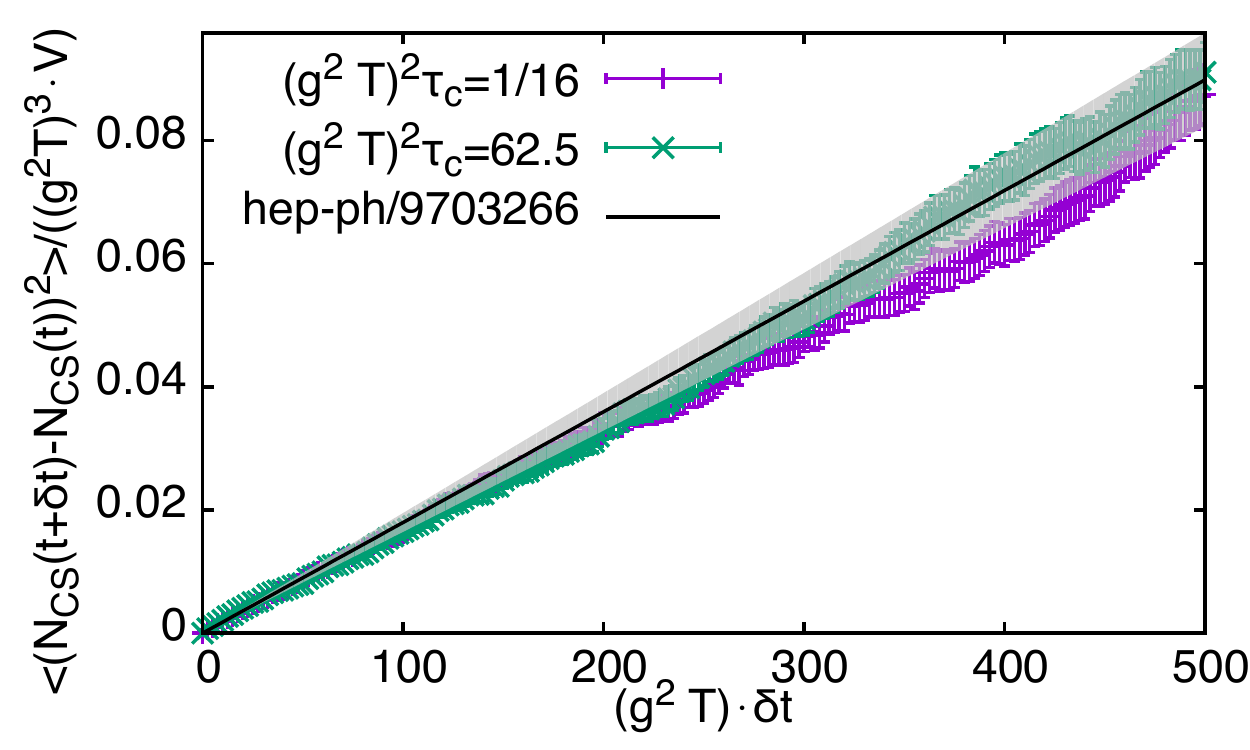}
\caption{Auto-correlation function of the Chern-Simons number for equilibrium configurations ($N=24$, $\beta=2$) with two different cooling depths $(g^2T)^2 \tau_{c}=0.0625$ and $62.5$. Our results are compared to the sphaleron rate from  \cite{Moore:1997cr}.}
\label{thermalcomparison}
\end{figure}

We now compute the auto-correlation function of the Chern-Simons number,
\begin{eqnarray}
\label{eq:NCSautocorr}
C(t,\delta t)=\frac{1}{V}\Big \langle \big( N_{CS}(t+\delta t) - N_{CS}(t) \big)^2  \Big\rangle\,,
\end{eqnarray}
which, according to Eq.~(\ref{rate}), can be used to define the sphaleron transition rate in the late time limit of the correlation function. Our results are shown in Fig.~\ref{thermalcomparison}, where we present the auto-correlation function in Eq.~(\ref{eq:NCSautocorr}) as a function of the temporal separation $\delta t$. We find that the auto-correlation function is approximately independent of the cooling depth $\tau_{c}$ indicating once again the dominance of topological transitions for the long-time behavior of the Chern-Simons number. One also observes from Fig.~\ref{thermalcomparison} that the $N_{CS}$ auto-correlation function shows an approximately linear rise as a function of $\delta t$. This result is consistent with the expectation that consecutive sphaleron transitions have the Markovian property of being uncorrelated with each other on sufficiently long time scales. The long time behavior of the Chern-Simons number can therefore be approximated by an integer random walk with the diffusion constant given by the sphaleron transition rate,
\begin{eqnarray}
\frac{1}{V} \Big\langle \Delta N_{CS}^{2}(\delta t) \Big\rangle _{eq} = \Gamma_{sph}^{eq}\, \delta t\;.
\end{eqnarray}
We compared our results for the sphaleron transition rate to the previous extraction by Moore \cite{Moore:1997cr}. The latter is represented by the black line with grey error bands in Fig.~\ref{thermalcomparison}. Excellent agreement is obtained between the two calculations. 

\begin{table}
\begin{tabular}{||c|c|c|c|c||}
\hline
$N$ & $\beta_{lat}$ & $g^2Ta$ & $(g^2T)^2~\tau_{c}$ & $N_{confs}$ \\
\hline
24 & 2 &  1/2 & 0.0625 & 1152 \\
24 & 2 &  1/2 & 4.0 & 196 \\
24 & 2 &  1/2 & 62.5 & 1180 \\
\hline
\end{tabular}
\caption{Data sets for studying sphaleron transitions in 3D $SU(2)$ Yang-Mills theory in thermal equilibrium.}
\label{tab:EqData}
\end{table}

\section{Sphaleron transitions in the Glasma}
Now that we have demonstrated that we can reproduce established results in the literature for computing the sphaleron transition rate in non-Abelian plasmas in thermal equilibrium, we can apply these techniques to explore topological transitions in the Glasma. We will begin with discussing the initial conditions for the Glasma in weak coupling. We will for simplicity not consider the realistic case of a Glasma undergoing rapid longitudinal expansion but will restrict ourselves to an isotropic system in a static box. In agreement with previous studies, we will show that even if one starts with a one scale problem given by the initial hard scale, the saturation scale $Q_s$, a scale separation develops with time between the hard scale and the softer electric and magnetic screening scales. Of particular interest is the temporal evolution of the spatial string tension, which we will compute for the first time. We will then compute the Chern-Simons number in the Glasma and study its auto-correlation behavior. Finally, we will demonstrate that one can express the extracted sphaleron transition rate in terms of the magnetic screening scale in the system. 

\label{nonequilibrium}
\subsection{Initial conditions \& Single particle spectra}
We choose our initial conditions to mimic the physical situation in a weak coupling scenario of an ultra-relativistic heavy ion collision at very high energies~\cite{Berges:2013fga,Berges:2013eia}. The defining feature of the non-equilibrium plasma, often called the Glasma, is an initial non-perturbatively large phase space occupancy $f(t_0,p)\sim 1/\alpha_S$ of gluon modes up to a saturation scale $Q_s$. As discussed previously, these initial conditions can be implemented in a simple quasi-particle picture where the initial gauge fields, and their momentum conjugate electric fields, are expressed as a superposition of transversely polarized gluons,
\begin{eqnarray}
A^a_\mu (t_0,x)&=& \sumint \frac{d^3 k}{(2 \pi )^2}\frac{1}{2k}\sqrt{f(t_0,k)} \left[ c_k^a \xi_\mu^{\lambda} (k)e^{ikx} + c.c.\right]\;,  \nonumber \\
E^a_\mu (t_0,x)&=& \sumint \frac{d^3 k}{(2 \pi )^2}\frac{1}{2k}\sqrt{f(t_0,k)} \left[ c_k^a \dot{\xi}_\mu^{\lambda} (k)e^{ikx} + c.c.\right], \nonumber \\
\end{eqnarray}
where $\xi_\mu^{\lambda} (k)$ labels the transverse polarization vectors, the $c\text{'s}$ are complex Gaussian random numbers with zero mean and unit variance and $ \sumint$ indicates a sum over polarizations and an integral over 
wavenumbers. The above initial conditions do not automatically satisfy the non-Abelian Gauss law $D_{\mu}E^{\mu}=0$; we must therefore enforce this constraint using the methods described in \cite{Moore:1996wn}. Our initial gluon distribution $f(t_0,k)$ is chosen to be
\begin{eqnarray}
f(t_0,k)= \frac{n_0}{2 \pi N_c \alpha_S} \frac{Q_s}{\sqrt{k^2+Q_s^2/10}} \theta(Q_s-k)\;,
\end{eqnarray}
where $n_0$ is a free parameter of order unity that can be used to vary the initial overoccupancy--if not stated otherwise, we choose $n_0=1$ as the default value.  

\begin{figure}
\centering
\includegraphics[width=0.5\textwidth,natwidth=610,natheight=642]{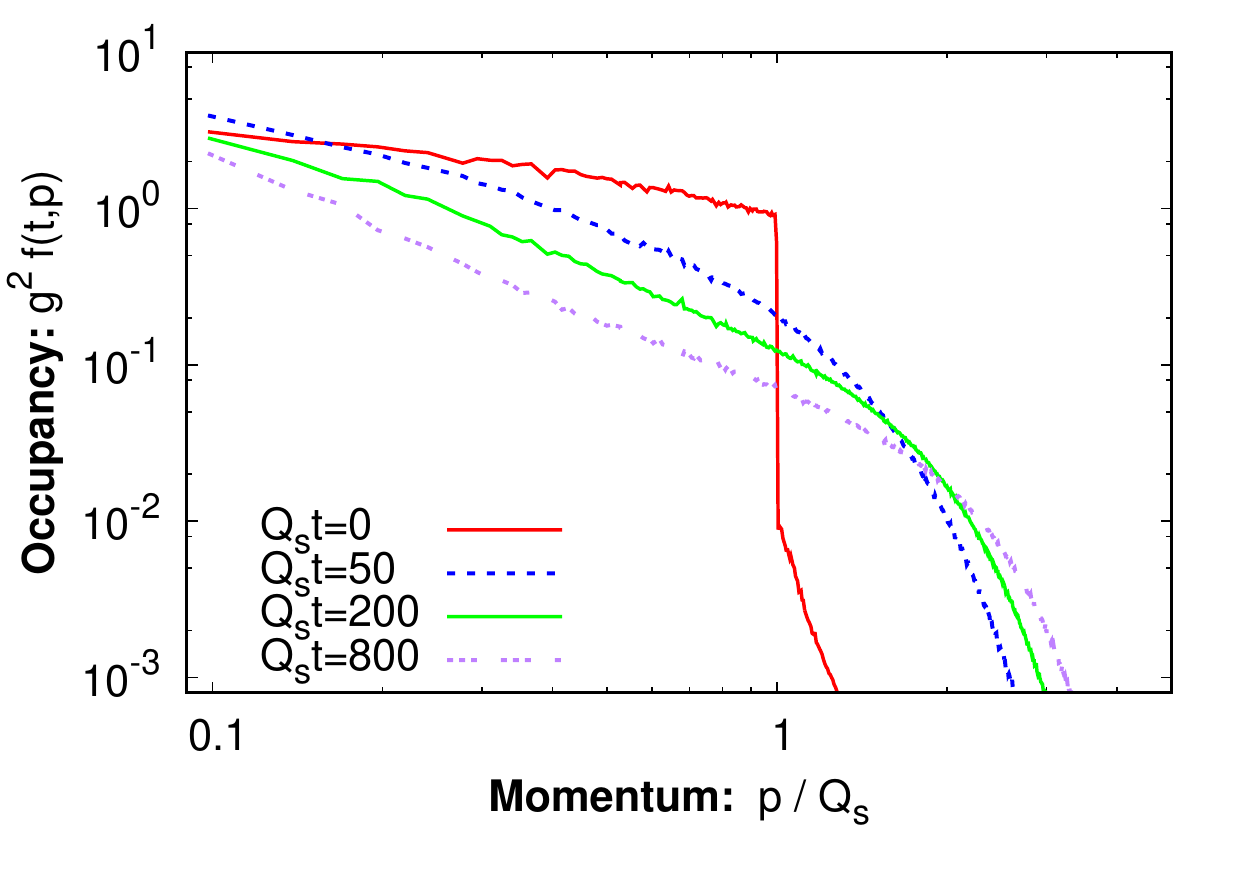}
\caption{Single particle spectra extracted at different times $Q_s t=0,50,200,800$ of the non-equilibrium evolution on a $N=128$ lattice with spacing $Q_s a=0.5$. }
\label{initialcondition}
\end{figure}

While our choice of initial conditions may appear peculiar at first sight, previous studies \cite{Berges:2013fga,Schlichting:2012es,Kurkela:2012hp} have demonstrated that the details of the initial conditions become irrelevant on a time scale $Q_s t \sim n_{0}^{-2}$, which corresponds parametrically to the inverse of the large angle scattering rate. Indeed, following the time evolution of the gluon spectrum\footnote{We extract the single particle spectrum from equal time correlation functions in Coulomb gauge as discussed in detail in \cite{Berges:2013fga}.} shown in Fig.~\ref{initialcondition}, one observes a rapid change of the gluon distribution at early times $Q_st \lesssim 50$. As first reported in \cite{Schlichting:2012es,Kurkela:2012hp}, the Glasma subsequently approaches a universal attractor solution characterized by an infrared power law with a rapid fall--off at high momenta. This evolution is shown for $Q_st \leq 800$ in Fig.~\ref{initialcondition}. In this regime, the dynamics is entirely characterized in terms of a self-similar scaling of the gluon distribution \cite{Berges:2013fga,Schlichting:2012es,Kurkela:2012hp}
\begin{eqnarray}
\label{eq:fScaling}
f(t,p)= (Q_s t)^{\alpha} f_{S}((Q_st)^{\beta}p)\;.
\end{eqnarray}
The scaling exponents $\alpha=-4/7$ and $\beta=-1/7$ in this regime have been extracted to high accuracy from classical-statistical simulations \cite{Berges:2013fga,Schlichting:2012es,Kurkela:2012hp} and can be understood from simple considerations in kinetic theory~\cite{Kurkela:2011ti,Blaizot:2011xf}. As discussed in \cite{Baier:2000sb,Berges:2013fga,} the self-similar scaling in Eq.~(\ref{eq:fScaling}) persists up to a timescale $Q_s t_{\text{Quantum}}=\alpha_s^{-7/4}$, when the typical occupancy of hard modes becomes of order unity. Beyond $t_{\text{Quantum}}$ a classical-statistical description is no longer accurate, as instead quantum effects become important and drive the system to thermal equilibrium \cite{Kurkela:2014tea}.

\subsection{Evolution of characteristic scales}
\label{scalemeasurements}
In a weakly coupled plasma in thermal equilibrium, the three scales are parametrically separated by powers of the coupling constant, with the hard scale $\sim T$ much larger than the electric screening scale $\sim g T$, much larger than the magnetic screening scale $\sim g^2T$.  However in the Glasma, initially all scales are on the order of the saturation scale $Q_s$ because of the non-perturbatively large occupancies $f\sim 1/\alpha_{S}$. Hence the hierarchy of scales characteristic of a weakly coupled plasma in equilibrium has to be developed dynamically during the thermalization process. 

Even though the existence of a scale hierarchy is essential for the applicability of most weak coupling methods, and plays a crucial role for the theoretical understanding of the sphaleron transition rate in thermal equilibrium, we emphasize that the classical-statistical lattice approach does not explicitly rely on a separation of scales. We will therefore use this approach in the following to study the time evolution of the characteristic scales starting from the non-perturbative high occupancy regime.

\begin{figure}
\centering
\includegraphics[width=0.5\textwidth,natwidth=610,natheight=642]{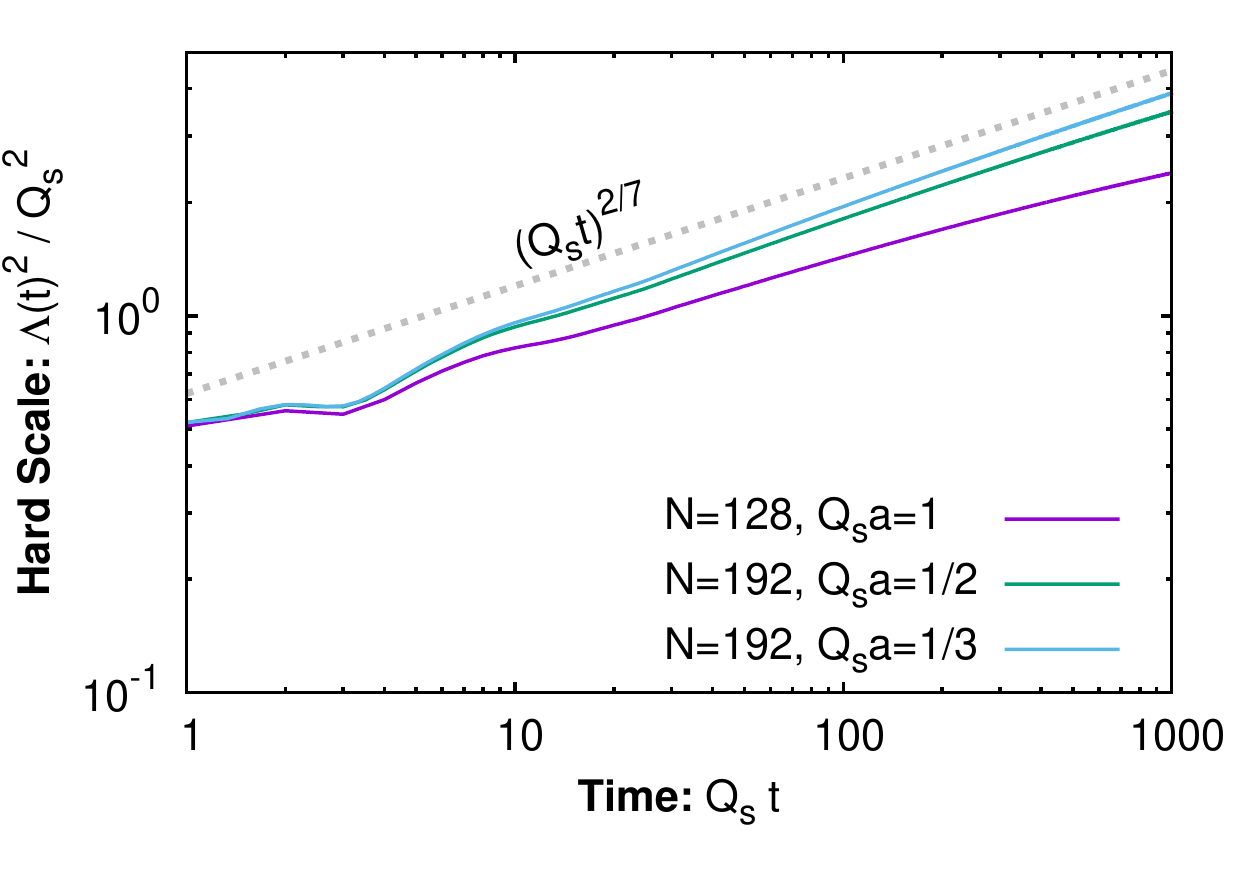}
\caption{Evolution of the hard scale $\Lambda^2(t)$ for three different lattice spacings $Q_s a=1,1/2,1/3$. The grey dashed line indicates a $(Q_st)^{2/7}$ scaling behavior.}
\label{hardscales}
\end{figure}

\subsubsection{Hard scale $\Lambda$}
We first consider the time evolution of hard scale $\Lambda(t)$, describing the characteristic momentum scale of hard excitations in the plasma. Following the methodology of previous Yang-Mills studies \cite{Kurkela:2012hp,Berges:2013fga,Berges:2013eia}, we determine this scale using the gauge invariant local operator definition
\begin{eqnarray}
\Lambda^2(t)=\frac{\langle H(t)\rangle}{\langle \epsilon_{B}(t) \rangle }\,,
\label{hardscalesquared}
\end{eqnarray}
where the dimension six operator $H(t)$ corresponds to the trace of the squared covariant derivative of the field strength tensor
\begin{eqnarray}
H(t)= \frac{1}{3V}\int d^3x~D^{ab}_{j}(x) F^{ji}_{b} (x) ~D^{ac}_{k}(x) F^{ki}_{c} (x)\,.
\end{eqnarray}
Here $\epsilon_{B}(t)$ denotes the magnetic energy density
\begin{eqnarray}
\epsilon_{B}(t)=\frac{1}{4 V} \int d^3x~ F_{ij}^{a}(x)  F_{ij}^{a}(x)\,.
\end{eqnarray}
Summation over the color $a,b,c=1,\cdots,N_c^2-1$ and spatial Lorentz indices $i,j,k=x,y,z$ is implied.  This hard scale can be expressed in perturbation theory as the ratio of moments of the single particle distribution as (see for example ~\cite{Berges:2013fga})
\begin{eqnarray}
\Lambda^2(t)= \frac{2}{3} \frac{\int d^3p~p^3~f(t,p) }{\int d^3p~p~f(t,p)} 
\end{eqnarray}
such that for a weakly coupled plasma in thermal equilibrium one has
\begin{eqnarray}
\Lambda^{2}_{eq}=\frac{80}{63} \pi^2 T^2\;.
\end{eqnarray}
whereas initially
\begin{eqnarray}
\Lambda^{2}_{init}= c_{\Lambda}~Q_s^2
\label{eq:hard-init}
\end{eqnarray}
where $c_{\Lambda}=\frac{236-6 \sqrt{11}}{525}$  up to higher order corrections for our choice of initial condition. Our results for the non-equilibrium evolution of the hard scale are shown in Fig.~\ref{hardscales} for three different lattice spacing $Q_s a=1,1/2,1/3$. Since the operator definition in Eq.~(\ref{hardscalesquared}) involves an ultraviolet sensitive dimension six operator, the result is only slowly convergent as a function of the lattice spacing. Moreover, since the physical hard scale $\Lambda$ increase as a function of time, clear deviations from the continuum limit can be observed for coarser lattices at late times. However, for the finer lattices we find that the time evolution of of the hard scale converges towards a $\Lambda^2 \sim (Q_st)^{2/7}$ scaling behavior as reported previously in \cite{Kurkela:2011ti,Berges:2013fga}. Indeed, this scaling is expected from the self-similar evolution of the single particle distribution in Eq.~(\ref{eq:fScaling}).

\begin{figure}
\centering
\includegraphics[width=0.5\textwidth,natwidth=610,natheight=642]{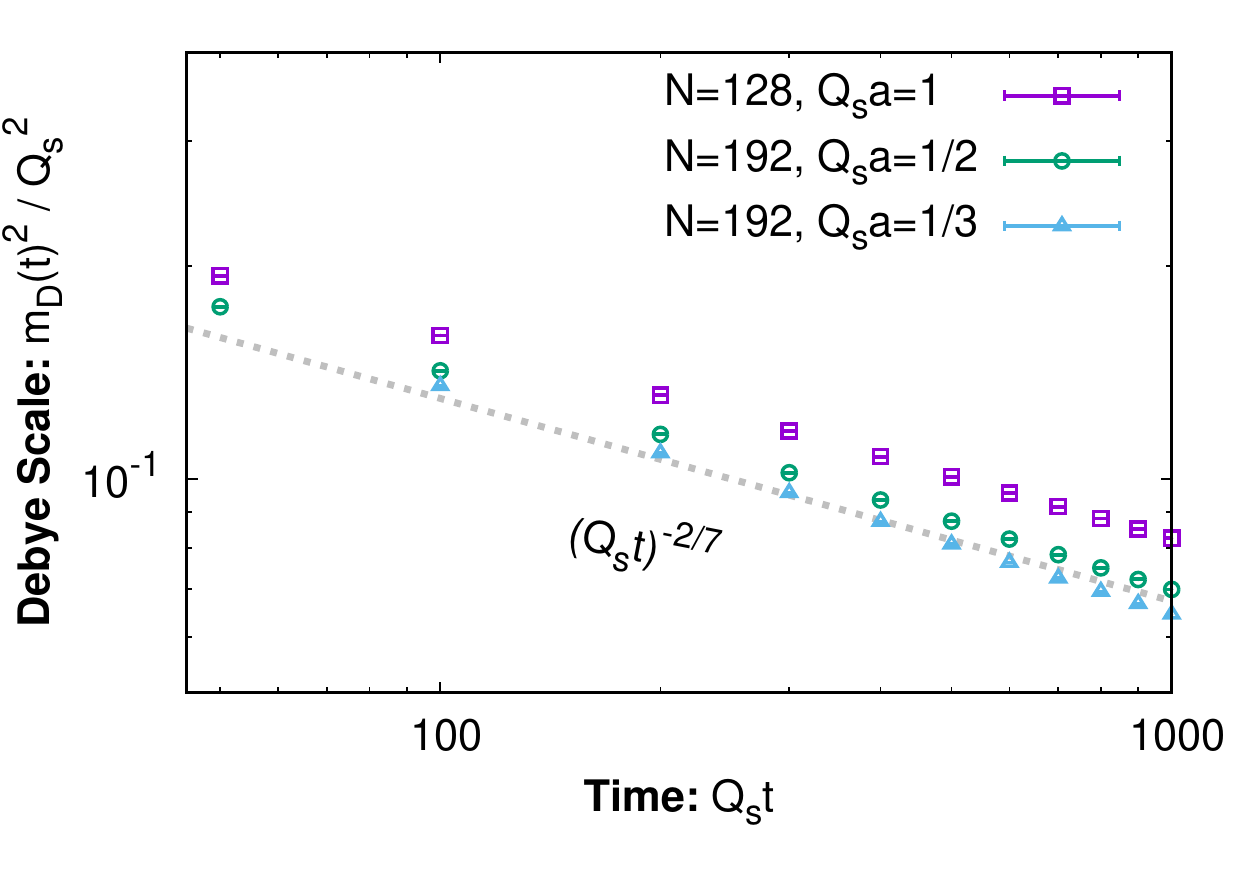}
\caption{Evolution of the Debye scale $m_{D}^2(t)$ for three different lattice spacings $Q_s a=1,1/2,1/3$. The grey dashed line indicates a $(Q_st)^{-2/7}$ scaling behavior.}
\label{debyemasssquared}
\end{figure}

\subsubsection{Debye scale $m_D$}
We will now discuss the extraction of the Debye mass, which corresponds to the (static) electric screening scale in the plasma. Since we are unaware of a non-perturbative definition of the Debye mass in the context of classical-statistical real-time lattice simulations\footnote{For a discussion of the non-perturbative definition of the Debye mass in Euclidean lattice gauge theories, see for instance \cite{Arnold:1995bh}. However it is not obvious how to adapt these concepts to our real time lattice simulations.} we follow previous works \cite{Kurkela:2011ti,Berges:2013fga} and instead use the leading order perturbative definition of the Debye screening mass 
\begin{eqnarray}
m_D^2(t)=4\;g^2N_c \int \frac{d^3p}{(2 \pi)^3} \frac{f(t,p)}{p}\;.
\label{debyemass}
\end{eqnarray}
The corresponding lattice expression is obtained by replacing the momentum integral $\int \frac{d^3p}{(2 \pi)^3}$  in this expression by the lattice sum over discrete momentum modes $\frac{1}{(N a)^3} \sum_p$. In a weakly coupled plasma in thermal equilibrium,
\begin{eqnarray}
m_{D,eq}^{2}=\frac{g^2 N_c T^2}{3}\,,
\end{eqnarray}
with the Debye scale is parametrically smaller than the typical hard momentum scale. In contrast, one finds initially in our non-equilibrium simulation that
\begin{eqnarray}
m_{D,init}^{2}= c_{m_{D}^2} n_0 Q_s^2\,.
\end{eqnarray}
where $c_{m_{D}^2} =\sqrt{\frac{2}{5}} \frac{\sqrt{11}-1}{\pi^2}$ for our choice of initial condition. Comparing this expression to Eq.~(\ref{eq:hard-init}). we observe that there is no parametric separation of scales at the initial time even at arbitrarily weak coupling. Our results for the non-equilibrium temporal evolution of the Debye scale $m_{D}^2(t)$ are depicted in Fig.~\ref{debyemasssquared}  for different lattice discretizations. We find that $m_D^2$ approaches an approximate $(Q_s t)^{-2/7}$ scaling behavior in the continuum limit, consistent with the expectation suggested by the scaling of the single particle distribution. 

\subsubsection{Spatial string tension $\sigma$}
\label{string-tension}
In thermal equilibrium, the sphaleron transition rate is controlled by the dynamics of modes with momenta on the order of the inverse magnetic screening length. It is therefore desirable to extract an equivalent quantity in our non-equilibrium setup and study its evolution in time. As in thermal equilibrium, one can investigate the behavior of spatial Wilson loop 
\begin{eqnarray}
\label{eq:WilsonLoop}
\mathrm{W}(t,A) = \mathcal{P} \exp \Big(i g \oint dx^{i} A_{i}(x,t)\Big)\,,
\end{eqnarray}
as a function of the area $A$ enclosed by the loop. Since the large distance behavior of the Wilson loop in thermal equilibrium is characterized by an area law,
\begin{eqnarray}
\label{eq:arealaw}
\Big\langle \text{Tr}~\mathrm{W}(A) \Big\rangle_{eq} \propto e^{-\sigma A}\,,
\end{eqnarray}
with the spatial string tension $\sigma$ related to the magnetic screening scale parametrically as
\begin{eqnarray}
\sigma \sim g^4 T^2\,,
\end{eqnarray}
at weak coupling, we can similarly extract the spatial string tension $\sigma$ in our non-equilibrium simulations and use $\sqrt{\sigma}$ as a proxy for the inverse magnetic screening length.

\begin{figure}
\centering
\includegraphics[width=0.5\textwidth,natwidth=610,natheight=642]{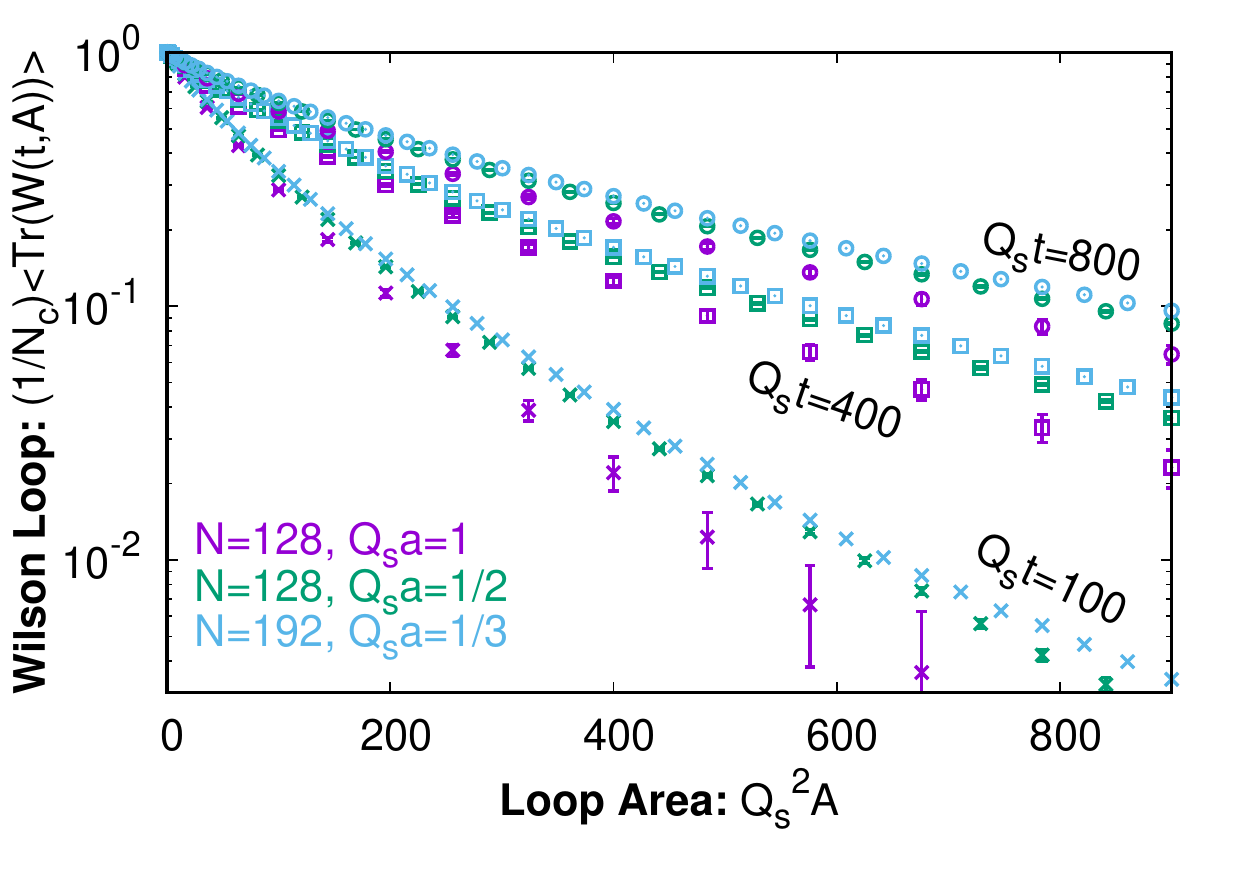}
\caption{Expectation value of the trace of the spatial Wilson loop versus the area of the loop at times $Q_s t=100,~400,~800$ (bottom to top). The different colored curves for each time correspond to different lattice discretizations.}
\label{wilsonlooptime}
\end{figure}

We first analyze the behavior of the spatial Wilson loop itself to establish the area law scaling in Eq.~(\ref{eq:arealaw}) for our non-equilibrium setup.\footnote{Our results presented in this section are obtained for so called on-axis Wilson loops, i.e. square loops with sides oriented along the lattice coordinate directions $(1,0,0),(0,1,0),(0,0,1)$. We have also investigated the behavior of so called off-axis Wilson loops where, instead of orienting the sides along the $(1,0,0),(0,1,0),(0,0,1)$ directions, the sides of the rectangle are oriented along any mutually orthogonal directions of lattice vectors (pointing for example in the $(2,1,0)$ and $(-1,2,0)$ directions). Within statistical errors the off-axis results agree with the on-axis measurements when expressed as a function of the area of the loop.} Our results for the (real-part of the) trace as a function of the area of the Wilson loop at three different times $Q_s t=100,~400,~800$ of the evolution are presented in Fig.~\ref{wilsonlooptime}. While for small areas the Wilson loops are close to the identity and do not exhibit scaling, convergence towards an area law scaling as in Eq.~(\ref{eq:arealaw}) can be observed for sufficiently large areas. Since the observation of such area law scaling in an out-of-equilibrium plasma is quite non-trivial, we briefly note that our results are in line with the findings reported in \cite{Berges:2007re,Dumitru:2014nka}.  Although area law scaling emerges clearly for all times shown in Fig.~\ref{wilsonlooptime}, extending our analysis to much earlier times is extremely challenging. This is because the values of the Wilson loop decrease rapidly as a function of area at very early times;  accurately resolving values $\lesssim 10^{-3}$ requires enormous statistics,  far exceeding the number of configurations $N_{confs}=4000$ used in Fig.~\ref{wilsonlooptime}.

\begin{figure}
\centering
\includegraphics[width=0.5\textwidth,natwidth=610,natheight=642]{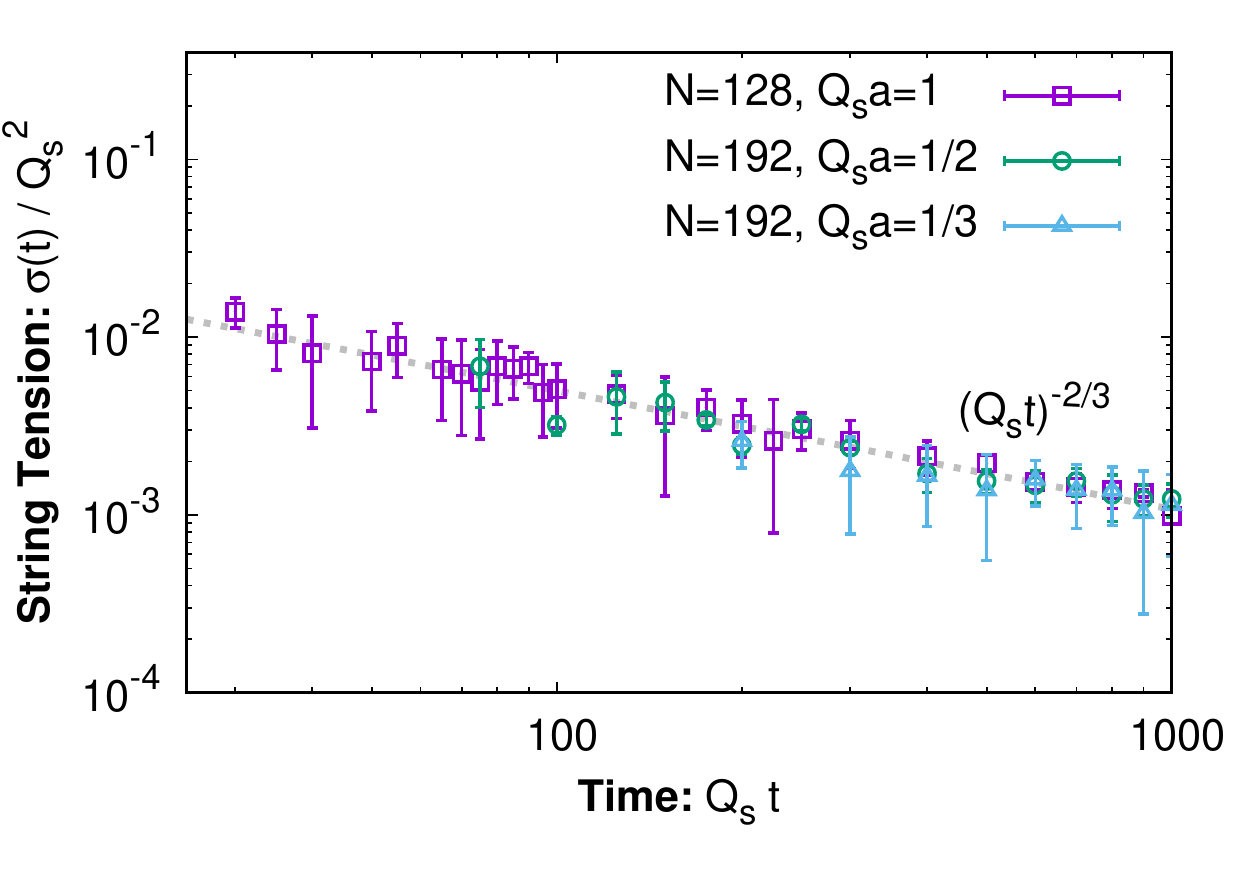}
\caption{Time evolution of the spatial string tension $\sigma(t)$ extracted from the logarithmic area derivative of the Wilson loop. }
\label{stringtension}
\end{figure}

We can further quantify the long distance behavior of the Wilson loop in terms of the spatial string tension $\sigma(t)$, which can be extracted from the logarithmic derivative of the Wilson loop with respect to the area
\begin{eqnarray}
\sigma(t)= - \left. \frac {\partial \ln\big\langle \text{Tr}~\mathrm{W}(t,A) \big\rangle}{ \partial A}\right|_{Q_s^2A \gg 1}\;.
\end{eqnarray} 
In practice, we compute the logarithmic derivative at different values of $Q_s^2A$ by performing a fit involving three adjacent data points. Subsequently, we search for an area independent value at $Q_s^2A \gg 1$ to extract the spatial string tension and include residual variations into our error estimate. 

Our results for the string tension $\sigma(t)$ as a function of time are shown in Fig.~\ref{stringtension}. Here we combined data for different lattice discretizations. While the error bars for the string tension are significantly larger relative to the previous measurements, one clearly observes a rapid decrease of the string tension as a function of time. We find that the time dependence can be approximately described by a $(Q_s t)^{-2/3}$ scaling behavior. We caution however that the precision with which we extract the exponent is limited to the 10 \% level -- primarily by the large statistical uncertainties in the measurement of the Wilson loop. We also note that in contrast to the time evolution of the hard and electric screening scales which can be estimated from kinetic theory~\cite{Blaizot:2011xf,Kurkela:2011ti}, we are not aware of an analytic prediction of the time evolution of the non-perturbative magnetic screening scale.
\begin{figure}
\centering
\includegraphics[width=0.5\textwidth,natwidth=610,natheight=642]{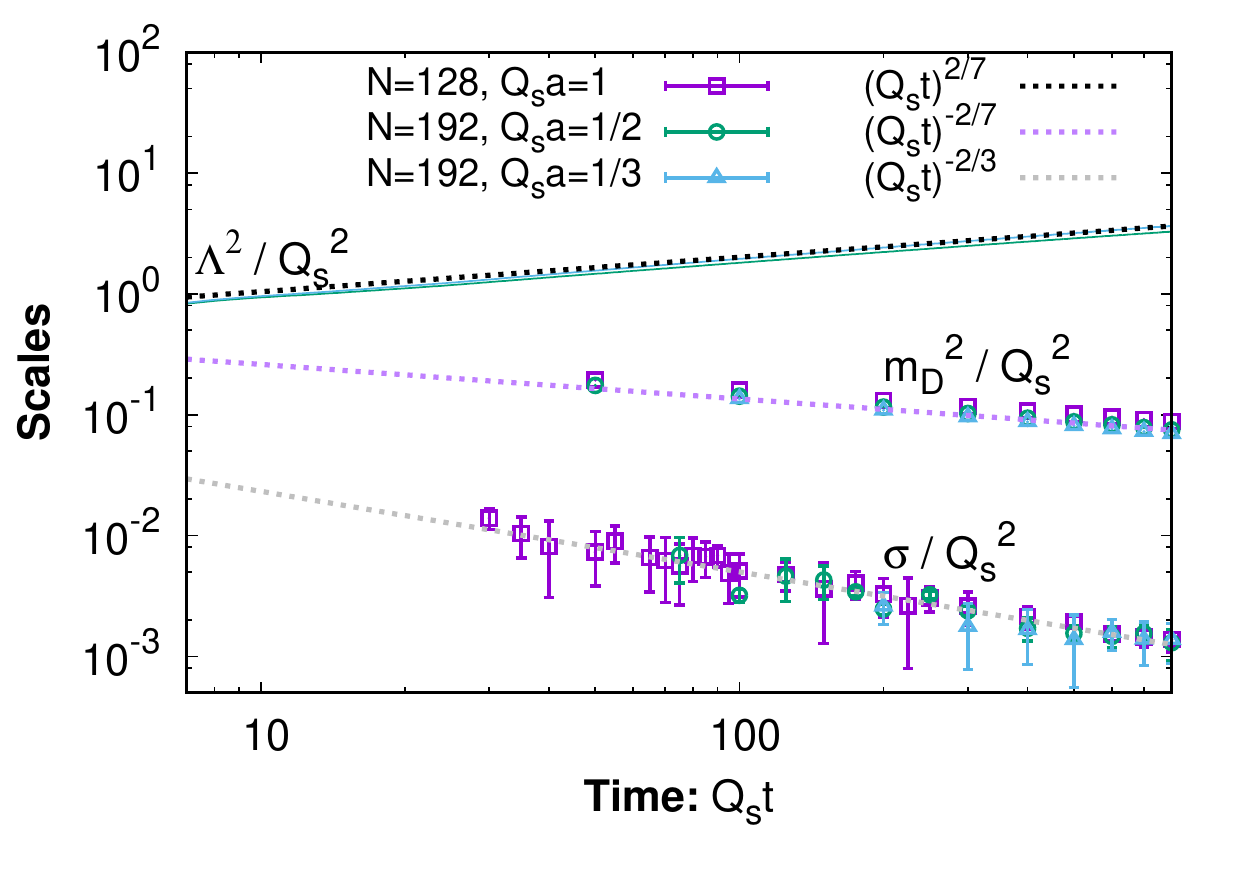}
\caption{Time evolution of the hard scale ($\Lambda^2$), the electric screening scale ($m_D^2$), and the spatial string tension ($\sigma$).  A clear separation of scales is established dynamically.}
\label{allscale}
\end{figure}

\subsubsection{Dynamical separation of scales}
Our results for the time evolution of the different characteristic scales of the Glasma are compactly summarized in Fig.~\ref{allscale}, where we plot the hard scale $\Lambda^2(t)$, the electric screening scale $m_{D}^{2}(t)$ and the spatial string tension $\sigma(t)$ (as a proxy for the magnetic screening scale) as a function of time. Initially all scales are of similar magnitude up to order one factors depending the details of the initial conditions.  Subsequently a clear separation of scales emerges dynamically as a function of time. While the hard scale increases according to a power law in time, the electric and magnetic screening scales decrease and separate from each other as well due to the faster decrease of the string tension relative to the Debye mass. Since such a separation of scales is essential to the applicability of standard (perturbative) weak coupling methods (such as effective kinetic descriptions or hard-loop effective field theories), our non-perturbative lattice results provide clear evidence that these methods become applicable at sufficiently late times.

\subsection{Sphaleron transitions \& evolution of Chern-Simons number} 
With the estimates obtained of the time evolution of the characteristic hard and soft scales in the Glasma, we will now proceed to study sphaleron transitions and the evolution of Chern-Simons number. We first note that a crucial difference from the thermal case is that we can simultaneously resolve the dynamics of hard as well as soft excitations using a classical-statistical lattice description. In the thermal case, the typical occupancies of hard modes $(p\sim T)$ are of order unity $f(p\sim T) \sim 1$ and a classical description does not apply to the hard modes. In contrast, in the Glasma, the hard modes $(p\sim\Lambda(t))$ are high occupied, $f(p\sim \Lambda(t)) \gg 1$, and therefore admit a classical description. Due to the noted complication, the study of sphaleron transitions in a weakly coupled plasma in thermal equilibrium proceeds via an effective field theory for soft $(p\sim g^2T)$ modes as discussed in Sec.~\ref{overview}. Since there is no such complication in the non-equilibrium case, we can directly study sphaleron dynamics in the Glasma using first principles lattice techniques.

\begin{table}
\begin{tabular}{||c|c|c|c|c||}
\hline
$N$ & $n_0$ & $Q_s a$ & $Q_s^2~\tau_{c}$ & $N_{confs}$ \\
\hline
64 & 0.5 &  1.0 & 12 & 256 \\
64 & 1.0 &  1.0 & 12 & 256 \\
64 & 1.0 &  1.0 & 324 & 1024 \\
64 & 1.5 &  1.0 & 12 & 256 \\
96 & 1.0 &  1.0 & 1 & 128 \\
96 & 1.0 &  1.0 & 12 & 4096 \\
96 & 1.0 &  1.0 & 36 & 1024 \\
96 & 1.0 &  1.0 & 108 & 1024 \\
96 & 1.0 &  1.0 & 216 & 1024 \\

\hline
\end{tabular}
\begin{tabular}{||c|c|c|c|c||}
\hline
$N$ & $n_0$ & $Q_s a$ & $Q_s^2~\tau_{c}$ & $N_{confs}$ \\
\hline
96 & 1.0 &  1.0 & 250 & 2048 \\
96 & 1.0 &  1.0 & 324 & 2048 \\
128 & 1.0 &  0.5 & 1 & 128 \\
128 & 1.0 &  0.5 & 12 & 512 \\
128 & 1.0 &  1.0 & 1 & 128 \\
128 & 1.0 &  1.0 & 12 & 2432 \\
192 & 1.0 &  0.5 & 12 & 512 \\
192 & 1.0 &  1.0 & 1 & 128 \\
192 & 1.0 &  1.0 & 12 & 512 \\
\hline
\end{tabular}
\caption{Data sets employed in the study of sphaleron transitions out-of-equilibrium.}
\label{tab:NonEqData}
\end{table}

This benefit is not without cost because for the Glasma all relevant scales have to be resolved simultaneously on the lattice. In addition, since the relevant scales are separating rapidly from each other with time, one typically requires very large lattices and the characteristic scales are accessible only for a limited amount of time.  A summary of the lattice parameters and data sets used in our study is provided in Tab.~\ref{tab:NonEqData}.   If not stated otherwise, all results shown were obtained for $N=96$ lattices with spacing $Q_sa=1$.

\subsubsection{Sphaleron transitions \& Chern-Simons number}
\begin{figure}
\centering
\includegraphics[width=0.5\textwidth,natwidth=610,natheight=642]{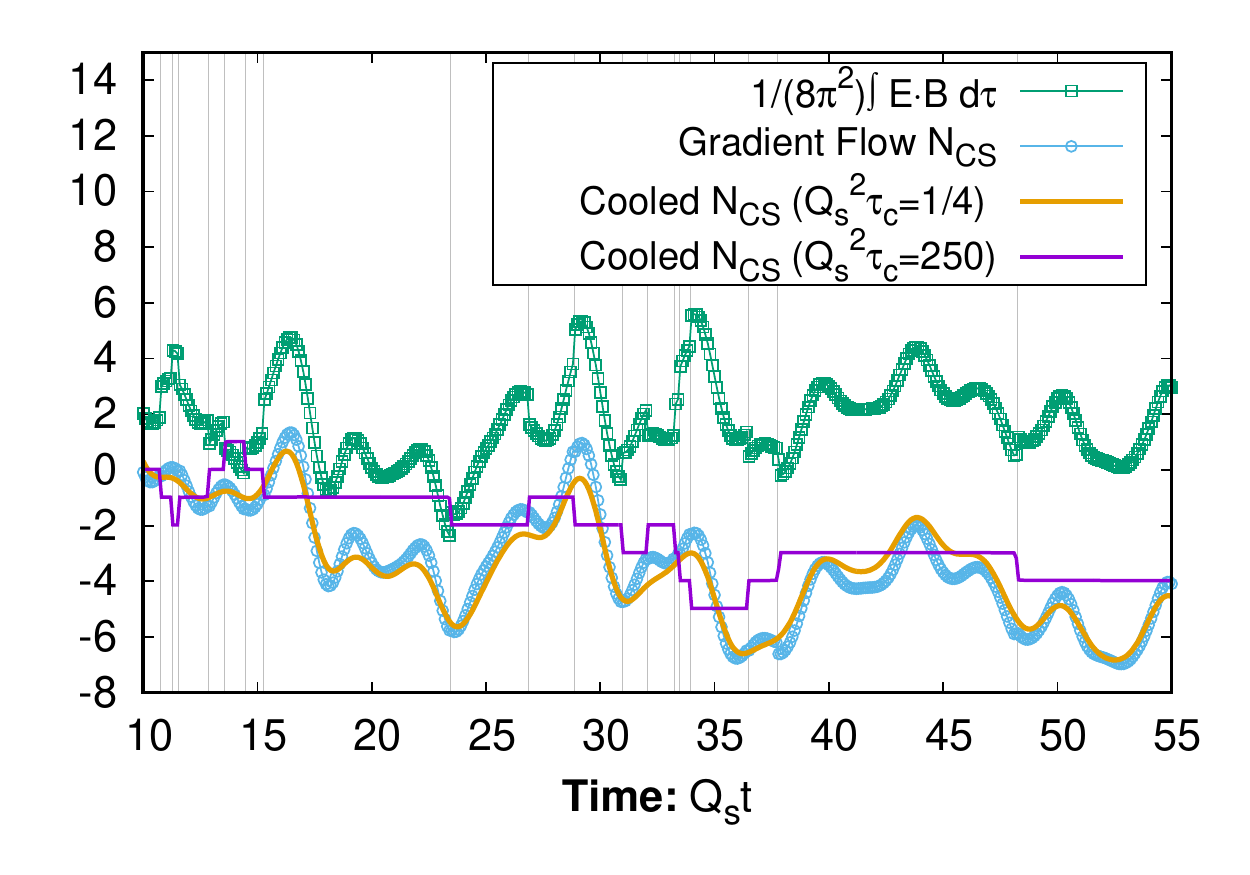}
\caption{Evolution of the Chern-Simons number for a single non-equilibrium configuration on a $N=96$ lattice with spacing $Q_sa=1$. Different curves correspond to different extraction procedures and contain variable amounts of field strength fluctuations in addition to the topological contributions.}
\label{discont}
\end{figure}

Since there are important conceptual differences of sphaleron dynamics in the Glasma relative to the thermal equilibrium case, we will begin by illustrating the time evolution of the Chern-Simons number and demonstrate our ability to successfully identify topological transitions. In Fig.~\ref{discont}, we show the time evolution of the Chern-Simons number for a short period of time during the non-equilibrium evolution of a single gauge field configuration. Different curves in Fig.~\ref{discont} correspond to different extraction methods and can be characterized as follows.

In the first case we perform gradient flow cooling of the non-equilibrium field configuration all the way to the vacuum and measure the integral of the Chern-Simons current along the cooling trajectory, as in Eq.~(\ref{eq:NCsVacCoolDef}). In this way we obtain the green curve in Fig.~\ref{discont}, which exhibits clear discontinuities. The positions of these discontinuities are indicated by the vertical gray lines. Each discontinuity corresponds to a transitions between different topological sectors defined by the so-called ``gradient flow separatrix"~\cite{Moore:1998swa}, which occurs when the gauge field configuration evolves from the basin of attraction of one vacuum state to the basin of attraction of another topologically inequivalent vacuum state.

By adding the Chern-Simons number $N_{CS}^{vac}(t)$ of the corresponding vacuum state, as in Eq.~(\ref{eq:NCsVacCoolDef}), we obtain the blue curve in Fig.~\ref{discont} corresponding to the gradient flow definition $N_{CS}^{gf}$ of the Chern-Simons number. One observes that $N_{CS}^{gf}$ is a continuous function of time. In addition to the topological information, it also contains contributions from finite energy fluctuations of the chromo-electric and chromo-magnetic fields. 

We have also compared our results from the gradient flow definition of the Chern-Simons number with the ones obtained from the calibrated cooling technique. When choosing a small cooling depth $Q_s^2\tau_{c}=0.25$, we obtain the orange curve in Fig.~\ref{discont}, which closely follows the gradient flow definition. By choosing a larger value of $Q_s^2\tau_{c}=250$ for the cooling depth, we can practically remove all field strength fluctuations above the vacuum and restrict the measurement of the Chern-Simons number to its topological content. Indeed one observes from the purple curve in Fig.~\ref{discont}, that for $Q_s^2\tau_{c}=250$ the evolution is characterized by discontinuous transitions between different topological sectors, which nicely coincide with the crossings of the gradient flow separatrix. 

Even though we can separate the topological content of the Chern-Simons number from the contribution of field-strength fluctuations at different length scales by varying the cooling depth, it is not a priori obvious which contributions to the Chern-Simons number are most relevant to the physics of the chiral magnetic effect. We will therefore vary the amount of cooling in the following and present results for different values of the cooling depth.

\begin{figure}
\includegraphics[width=0.5\textwidth,natwidth=610,natheight=642]{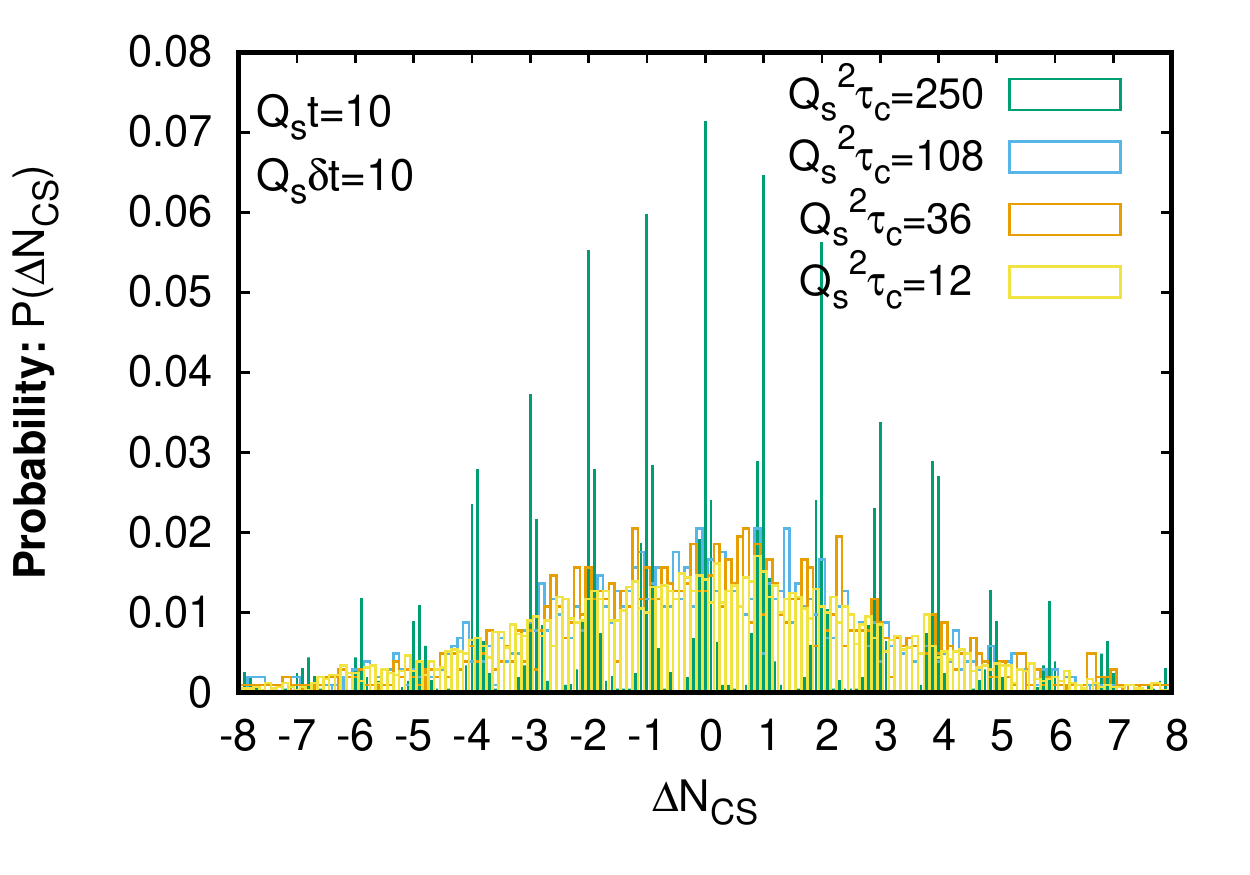}
\newline
\includegraphics[width=0.5\textwidth,natwidth=610,natheight=642]{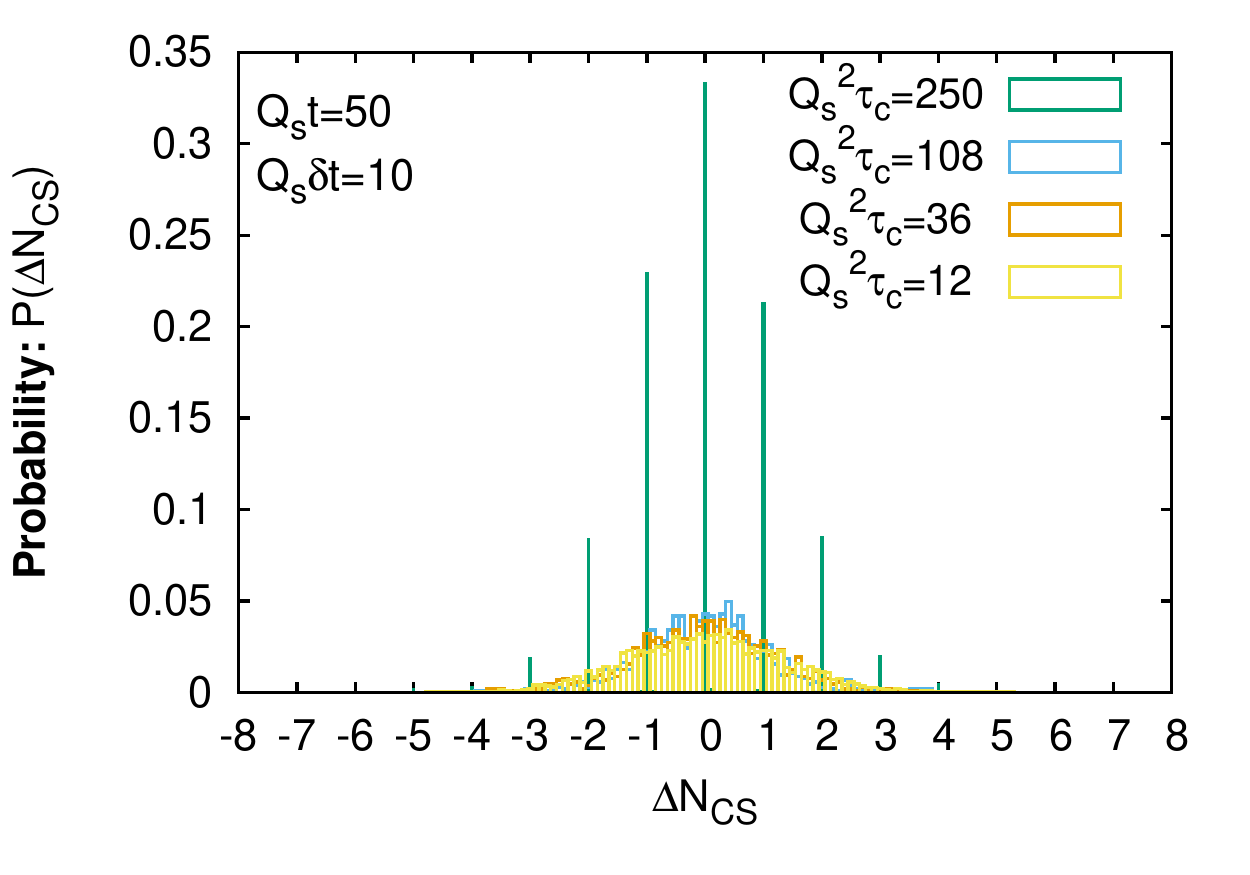}
\newline
\includegraphics[width=0.5\textwidth,natwidth=610,natheight=642]{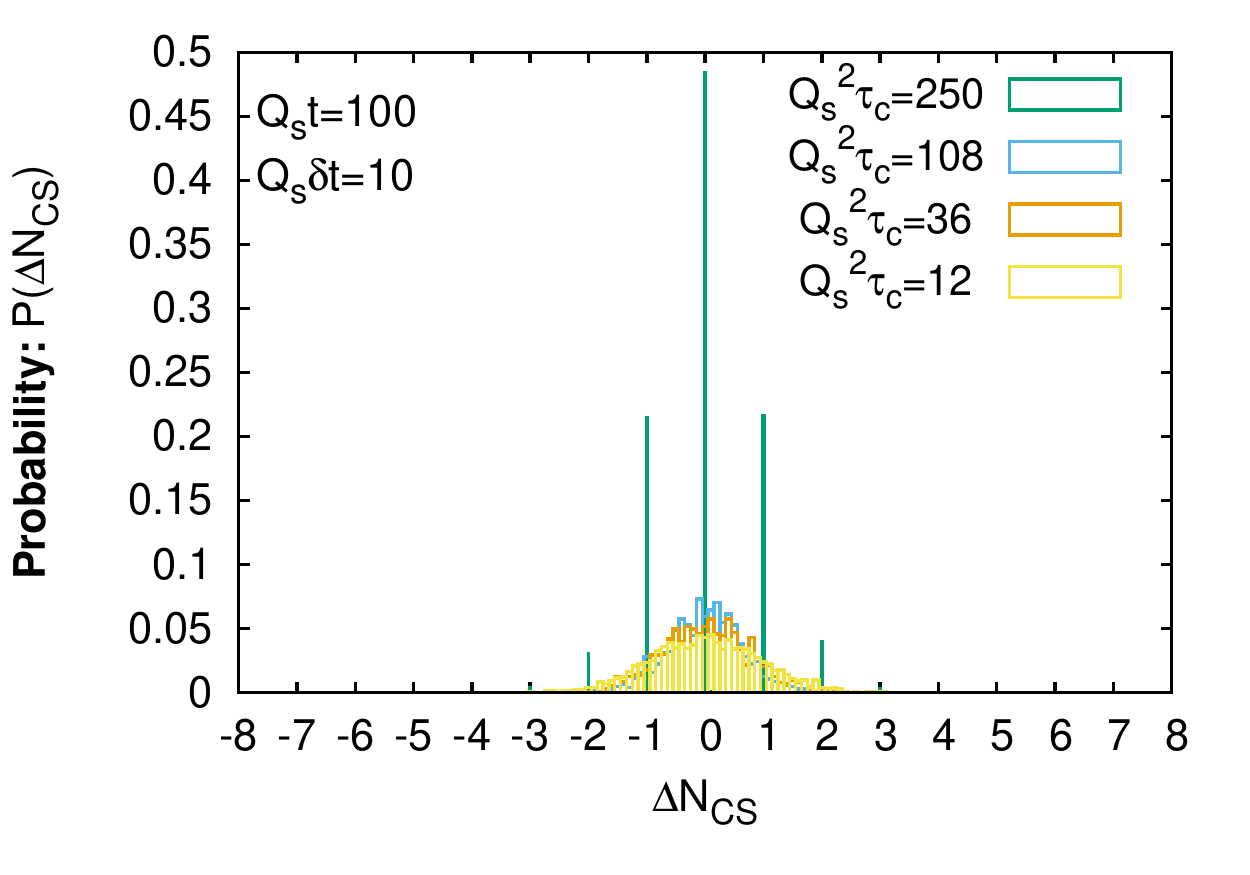}
\caption{Histograms of the distribution of the Chern-Simons number difference $\Delta N_{CS}$ within $Q_s\delta t=10$ units of time difference on a $N=96$, $Q_s a=1.0$ lattice. The different panels correspond to different reference times $Q_st=10,~50,~100$ (top to bottom) of the non-equilibrium evolution.}
\label{nonthermalhistograms}
\end{figure}

\subsubsection{Statistical analysis of Chern-Simons number}
Now that we have established that we are  able to identify topological transitions out-of-equilibrium, we will proceed with a more detailed statistical analysis. In order to obtain a first estimate of the time dependence of the transition rate, we follow the methodology in Sec.~\ref{sec:thermal} and first investigate the probability distribution of the difference $\Delta N_{CS}$ between the measurements of the Chern-Simons number at reference time $t$ and subsequent time $t+\delta t$. (See the  definition in Eq.~(\ref{eq:DeltaNCSDef}).) Our results for three different reference times $Q_s t=10,~50,~100$ during the non-equilibrium evolution and a common separation of $Q_s \delta t=10$ are shown in Fig.~\ref{nonthermalhistograms}. Irrespective of the amount of cooling, which primarily affects how narrowly the distributions are peaked around integer values, a clear difference between the different panels emerges. At early times of the non-equilibrium evolution, transitions between different topological sectors occur frequently resulting in a broad probability distribution for  $Q_st=10$.  At the later times $Q_st=50,~100$, the rate of transitions decreases rapidly as a function of time leading to much narrower distributions. Most prominently, starting at $Q_s t=100$ one finds that after a time $Q_s \delta t=10$ approximately half of the field configurations can still be found in the same topological sector. Most of the other half has merely transitioned to the neighboring topological sector. Very few configurations have $|\Delta N_{CS}| > 1$ when sufficient cooling $(Q_s^2\tau_{c}=250)$ is applied to isolate the topology at these late times.\footnote{We emphasize that the detailed fractions depend on the physical volume determined by the lattice size $V=(Na)^3$. Most importantly, we will demonstrate shortly that the variance $\langle \Delta N_{CS}^2 \rangle$ of the distributions exhibits the expected volume scaling.}

\begin{figure}
\centering
\includegraphics[width=0.5\textwidth,natwidth=610,natheight=642]{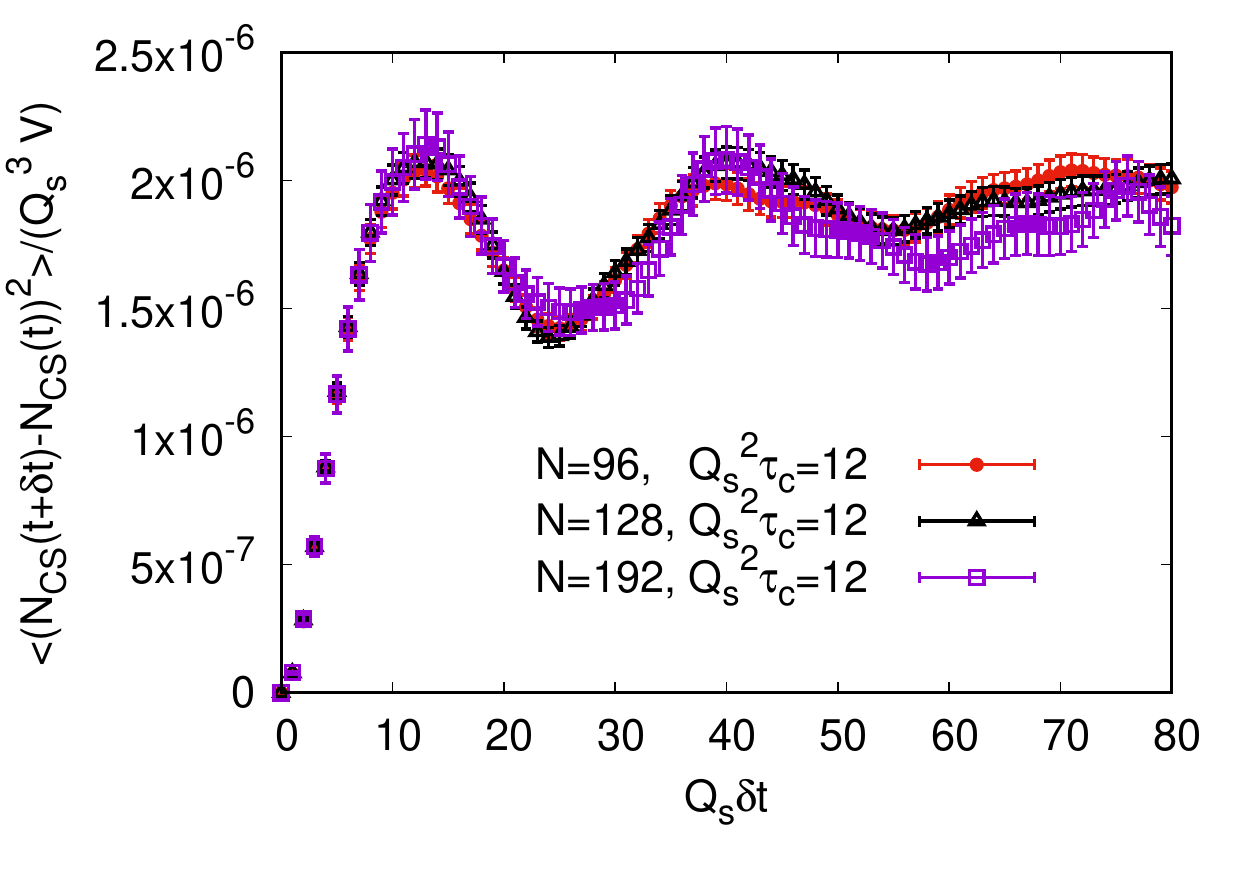}
\includegraphics[width=0.5\textwidth,natwidth=610,natheight=642]{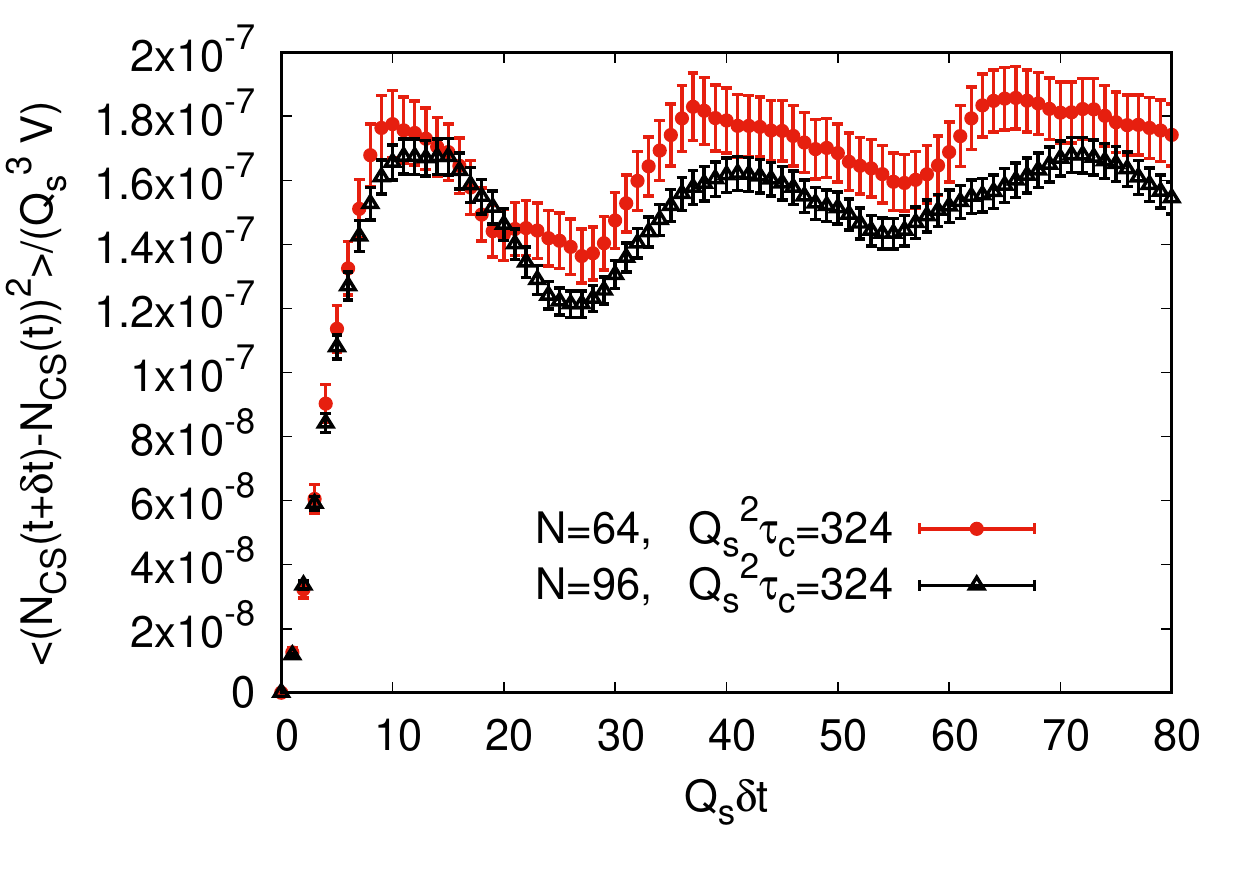}
\caption{Chern-Simons number auto-correlation function $ \frac{1}{Q_s^3 V} \langle (N_{CS}(t+\delta t)- N_{CS}(t))^2 \rangle$ as a function of the temporal separation $\delta t$ starting from $Q_s t=50$ during the non-equilibrium evolution. Different curves in each panel  correspond to different lattice volumes $Q_s^3 V=96^3,128^3,196^3$ for cooling depth $Q_s^2\tau_c=12$ (top) and $Q_s^3 V=64^3,96^3$ for  $Q_s^2\tau_c=324$ (bottom).}
\label{correlations}
\end{figure}

\begin{figure}
\centering
\includegraphics[width=0.5\textwidth,natwidth=610,natheight=642]{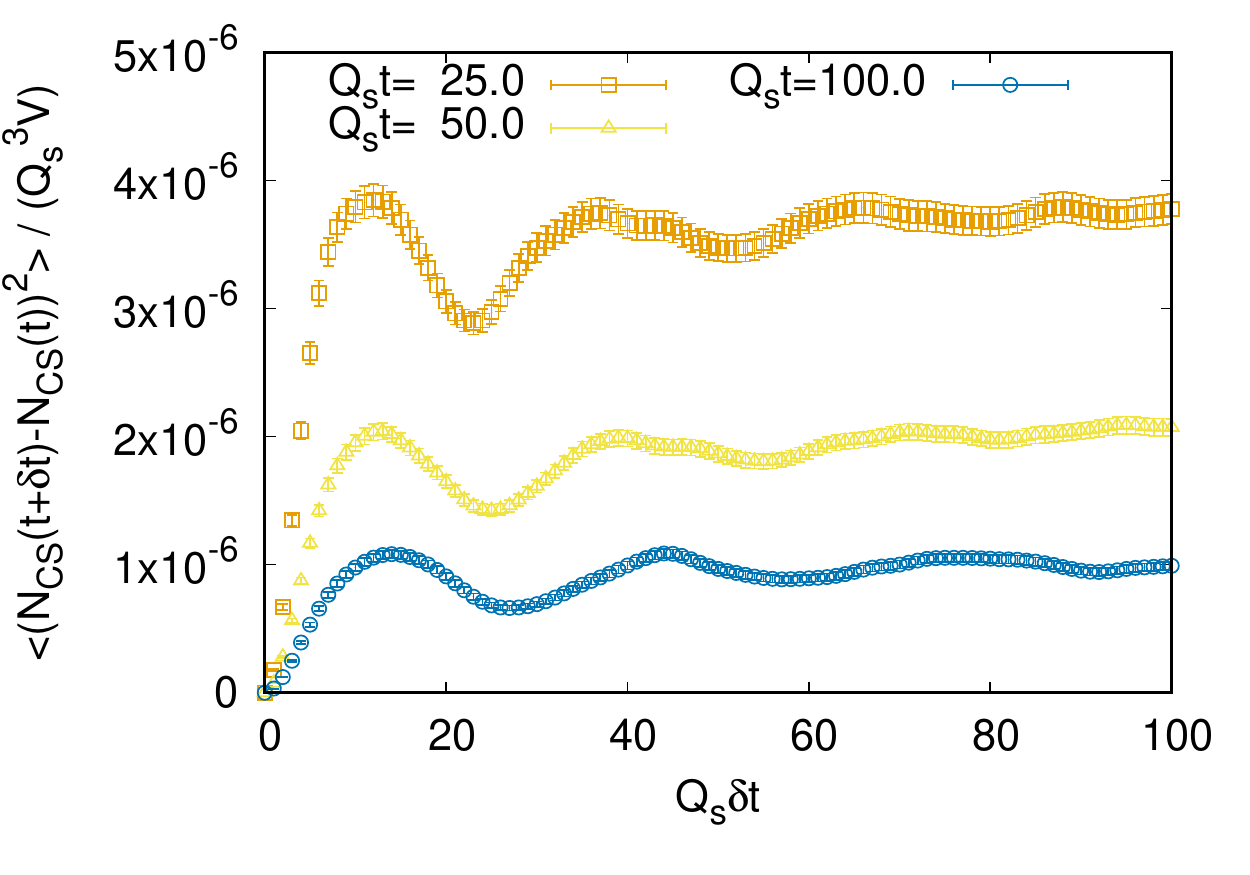}
\caption{Chern-Simons number auto-correlation function $ \frac{1}{Q_s^3 V} \langle (N_{CS}(t+\delta t)- N_{CS}(t))^2 \rangle$ as a function of the temporal separation $\delta t$, at three different times $Q_s t=25,~50,100$. Cooling depth is $Q_s^2 \tau_{c}=12$ in all cases.}
\label{correlatorattempts}
\end{figure}

While a rapid decrease of the sphaleron transition rate may be expected based on our previous observation that the magnetic screening scale decreases significantly as a function of time during the non-equilibrium evolution, we would like to further quantify this effect and determine the physical scales associated with the rate. We follow the same methodology devised in the equilibrium case and study the auto-correlation functions of the Chern-Simons number. Our results for the auto-correlation function of the Chern-Simons number $C(t,\delta t)$ are summarized in Figs.~\ref{correlations} and \ref{correlatorattempts}.

Fig.~\ref{correlations} shows results for the auto-correlation function for a fixed reference time $Q_s t=50$ during the non-equilibrium evolution. Starting with a rapid rise for small separations $Q_s\delta t \lesssim 10$, the growth of $\langle \Delta N_{CS}^2 \rangle$ slows down dramatically at larger $\delta t$ and is superseded by pronounced oscillations. In fact the oscillations are so significant that for certain periods such as e.g. $Q_s \delta t \simeq 10-20$  the variance $\langle \Delta N_{CS}^2 \rangle$ decreases with increasing separation $\delta t$. We have verified that this oscillatory behavior persists even for deeper cooling, as can be observed from the lower panel of Fig.~\ref{correlations} where we show results for $Q_s^2\tau_c=324$. Most importantly, this behavior is robust under variations of the lattice volume as can be observed from the different curves in Fig.~\ref{correlations} which agree with each other within errors -- indicating that our results are not significantly affected by finite volume effects. 

We emphasize that the non-monotonic behavior observed in our non-equilibrium simulations is clearly different from the thermal case, shown in Fig.~\ref{thermalcomparison}, where in contrast  $\langle \Delta N_{CS}^2 \rangle$ increases monotonously as a function of the separation time. Moreover the non-monotonic behavior observed in Fig.~\ref{correlations} is inconsistent with the usual interpretation of the Chern-Simons number evolution as a Markovian process. This is the case even when a time-dependent transition rate is assumed. It instead points to the fact that essential features of the dynamics on these time scales are non-Markovian as the evolution of the Chern-Simons number exhibits pronounced memory effects. 

While the non-monotonic behavior of the Chern-Simons number auto-correlation may come as a surprise, we find that it is a robust feature of the non-equilibrium evolution also at later times. This is seen in Fig.~\ref{correlatorattempts}, where we plot the Chern-Simons auto-correlation function starting from three different reference times $Q_s t=25,50,100$ during the non-equilibrium evolution. Even though a clear time dependence is seen in Fig.~\ref{correlatorattempts}, the general oscillation pattern remains intact albeit a slight change in the oscillation frequency can be observed. 

The most prominent feature of Fig.~\ref{correlatorattempts} though is the change in magnitude between different times $Q_st=25,~50,~100$. Clearly, the overall magnitude of $\langle \Delta N_{CS}^2 \rangle$ is significantly larger at earlier times and confirms our previous observation of a larger rate of topological transitions early on in the evolution of the Glasma. We will further quantify this statement below by extracting the rate of topological transitions associated with the initial rise of the auto-correlation function.

\subsection{Quantifying the rate of topological transitions}
We established clearly that Chern-Simons number evolution in the Glasma is non-Markovian. The equilibrium definition of the sphaleron transition rate in Eq.~(\ref{rate}) as the slope of the $N_{CS}$ auto-correlation function in the late time limit is therefore not sufficient to quantify topological transitions in the Glasma. However, for short enough separations in time $Q_s \delta t \lesssim 10$, the auto-correlation function does exhibit a rapid and approximately linear rise.  We will exploit this behavior to quantify the time dependence of the transition rate as 
\begin{eqnarray}
\label{eq:NonEqSTR}
\Gamma_{sph}^{neq}(t)=  \left\langle \frac{ (N_{CS}(t+\delta t)- N_{CS}(t))^2 } {V~\delta t} \right\rangle_{Q_s \delta t <10}\;.
\end{eqnarray}
This definition of the rate does not by any means single out topological contributions; it potentially receives large contributions from fluctuations of the color-electric and color-magnetic strength on all scales. We will therefore further apply different levels of cooling $\tau_c$ to efficiently suppress field-strength fluctuations on short distance scales. In particular, by choosing $Q_s^2 \tau_c \gg 1$ to cool almost all the way to the vacuum we can effectively suppress non-topological fluctuations and obtain a rate more closely related to the extraction of the sphaleron transition rate in thermal equilibrium\footnote{Similarly, we could also apply cooling to the vacuum and simply count the number of transitions defined by crossings of the gradient flow separatrix per unit time. However as some of the separatrix crossings do not affect the evolution on longer time scales, this measurement would differ by a ``dynamical prefactor" \cite{Moore:1998swa} which potentially also depends on time.}.

\begin{figure}
\centering
\includegraphics[width=0.5\textwidth,natwidth=610,natheight=642]{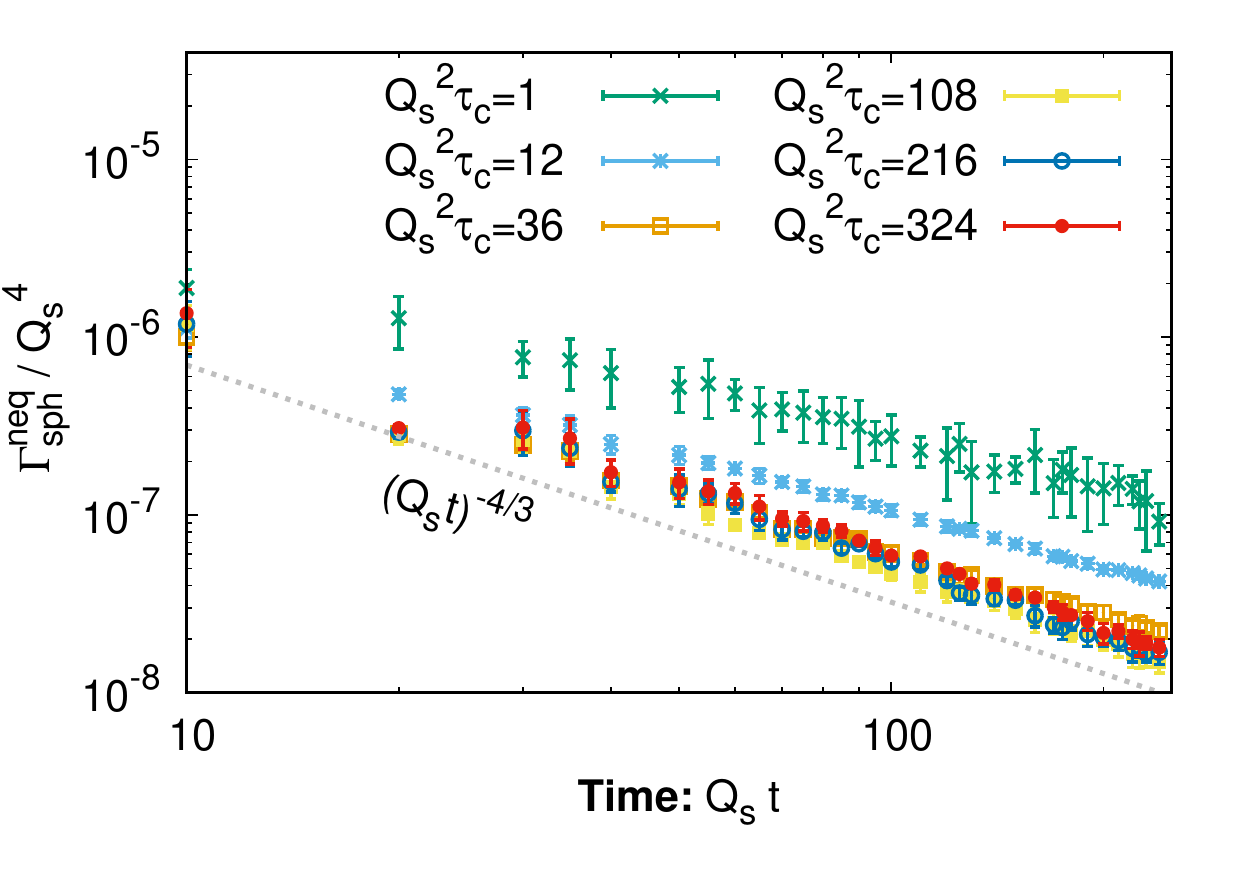}
\caption{Non-equilibrium sphaleron transition rate $\Gamma_{sph}^{neq}(t)$ as a function of time $Q_s t$ for various different values of the cooling depth $Q_s^2 \tau_c$.}
\label{therate}
\end{figure}

Our results for the non-equilibrium sphaleron transition rate $\Gamma_{sph}^{neq}(t)$ as a function of time are presented in Fig.~\ref{therate} for various levels of cooling. For small values of $Q_s^2 \tau_c \sim \mathcal{O}(1)$, a significant dependence on the cooling depth is observed.  This indicates large contributions to the rate from fluctuations of the color-electric and color-magnetic strength at short distance scales. In contrast, the results for large values of  $Q_s^2 \tau_c \gg 1$ appear to converge towards a  single curve isolating the genuine contributions due to topological transitions.

\begin{figure}[t!]
\centering
\includegraphics[width=0.5\textwidth,natwidth=610,natheight=642]{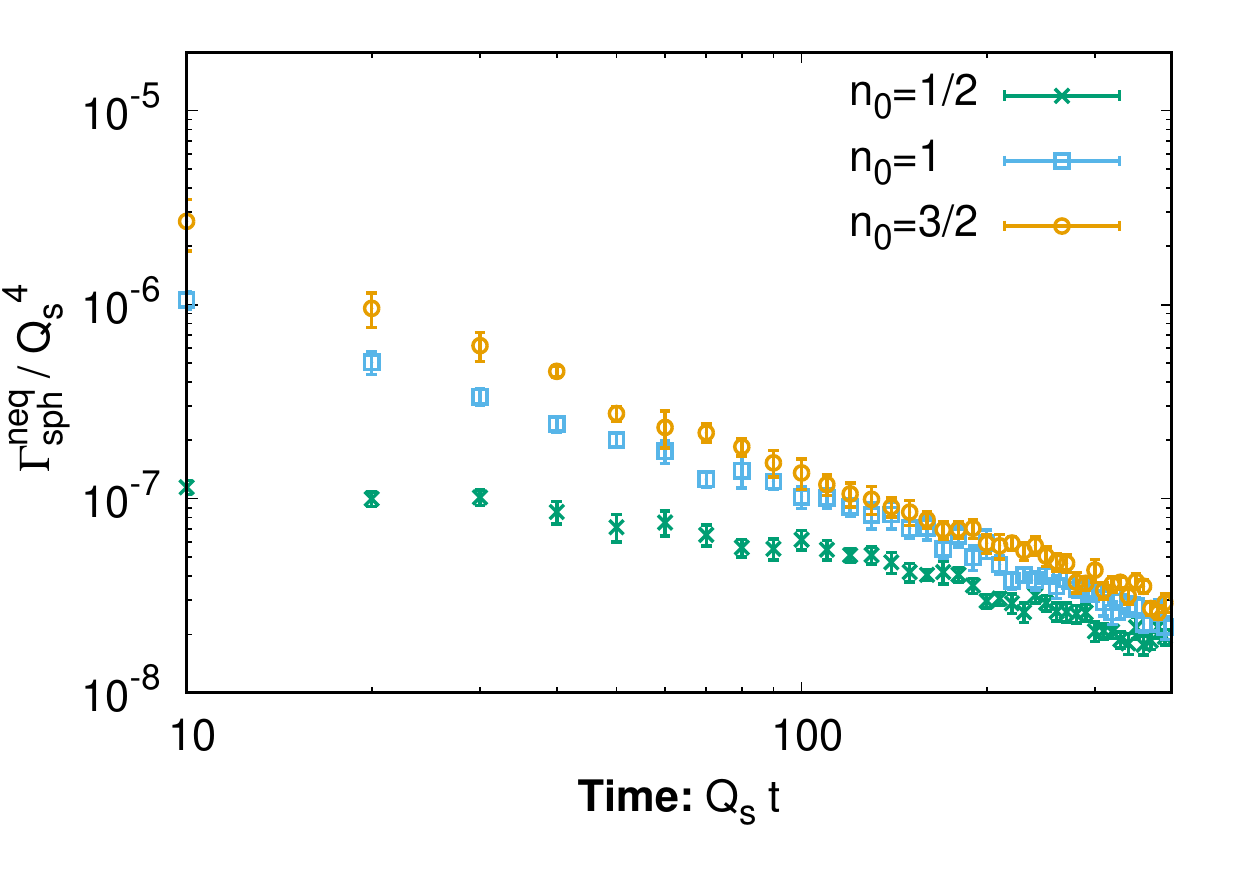}
\caption{Non-equilibrium sphaleron transition rate $\Gamma_{sph}^{neq}(t)$ as a function of time $Q_s t$ for different initial over-occupancies $n_{0}=3/2,~1,~1/2$. All curves have a cooling depth of $Q_s^2\tau_c=12$.}
\label{occupancy}
\end{figure}

Irrespective of the cooling depth, a clear time dependence of the rate can be observed. Starting from the largest values at early times, the rate rapidly decreases as a function of time and eventually approaches a power law behavior $(Q_s t)^{-\zeta}$ with an approximate scaling exponent $\zeta \simeq 4/3$. Albeit the early time behavior depends on the details of the initial conditions, we find that variations of the initial conditions do not affect the scaling behavior at later times. This is shown in  Fig.~\ref{occupancy} where the non-equilibrium sphaleron transition rate is plotted for different initial over-occupancies $n_{0}$; all the curves approach a common scaling behavior around $Q_s t \sim 100$. On the other hand, it is also clear that the sphaleron transition rate is by far the largest at early times and one should therefore expect a significant sensitivity to early time dynamics in time integrated quantities such as the axial charge density.

\begin{figure}[t!]
\centering
\includegraphics[width=0.5\textwidth,natwidth=610,natheight=642]{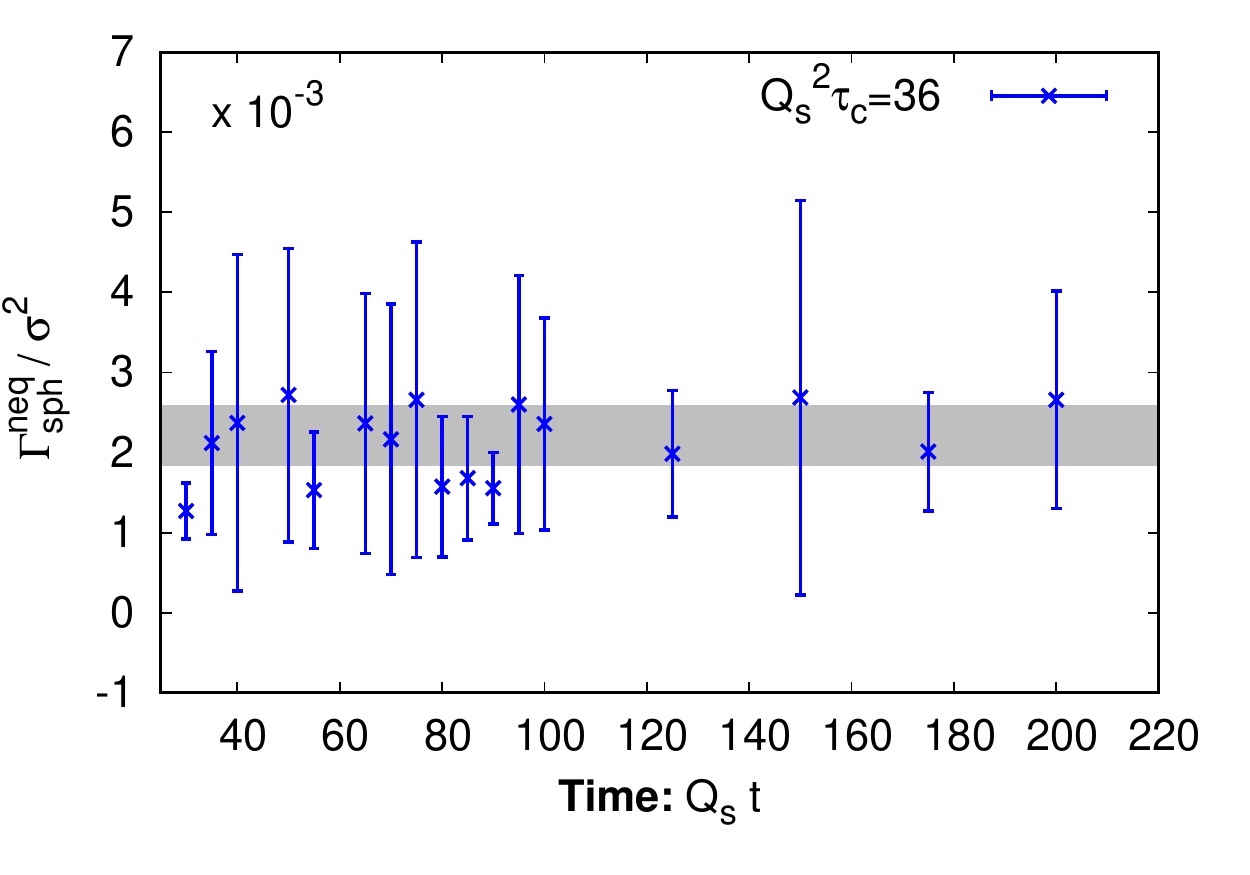}
\caption{Non-equilibrium sphaleron transition rate normalized by square of the spatial string tension.}
\label{gammaoversigma}
\end{figure}

The time dependence of the sphaleron transition rate in the scaling regime can be compared to those of the characteristic scales  of the Glasma. As we noted previously, the sphaleron transition rate in equilibrium is most sensitive to modes on the order of the magnetic screening scale.  One can analogously express the corresponding rate in the Glasma in units of the spatial string tension previously extracted in Sec.~\ref{string-tension}. Our results for the dimensionless ratio $\Gamma_{sph}^{neq}(t)/ \sigma^2(t)$ are presented in Fig.~\ref{gammaoversigma} as a function of time. While both the square of the spatial string tension as well as the non-equilibrium sphaleron rate change by an order of magnitude over the time scale shown in Fig.~\ref{gammaoversigma}, the ratio of the two quantities remains approximately constant with $\Gamma_{sph}^{neq}/ \sigma^2=(2.2 \pm 0.4) \cdot 10^{-3}$ extracted from the result shown in Fig.~\ref{gammaoversigma}. We interpret this result as clear evidence that the dynamics of sphaleron transitions off-equilibrium is fully determined by modes on the order of the inverse magnetic screening length in the Glasma.

\section{Conclusion \& Outlook}
\label{conclusion}
We presented a first study of the dynamics of sphaleron transitions in the Glasma -- the overoccupied and off-equilibrium non-Abelian plasma formed at early times in ultra-relativistic heavy ion collisions. For simplicity, we considered the Glasma dynamics for SU(2) gauge fields in a fixed box at the very weak couplings where classical-statistical dynamics captures its early time properties. 

The Glasma at the earliest time in our simulation is characterized by a single hard scale $Q_s$, defined as the  momentum up to which all modes in the Glasma have maximal occupancy of order $1/\alpha_S$; the occupancy falls sharply beyond $Q_s$. We showed that novel soft electric and magnetic scales develop in the Glasma and separate from each other and the hard scale dynamically with characteristic power laws in time that are independent of details of the initial conditions. Such a separation of scales is essential for the thermalization process since, in weak coupling, a clean hierarchy of these scales describes the equilibrium dynamics of a non-Abelian plasma. 

\begin{figure}
\centering
\includegraphics[width=0.5\textwidth,natwidth=610,natheight=642]{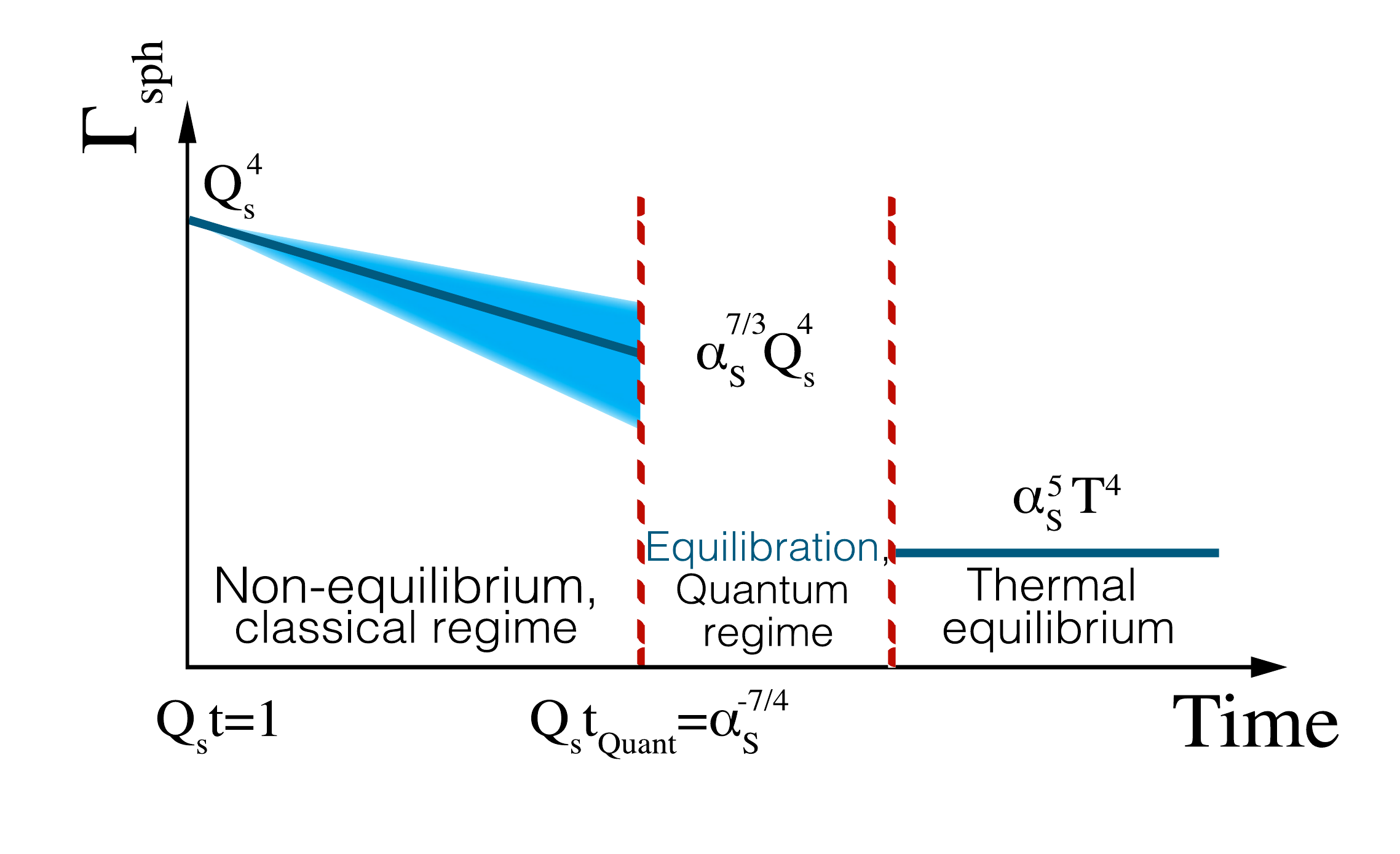}
\caption{Sketch of the temporal evolution of the sphaleron transition rate in the non-expanding Glasma. Shaded band represents the uncertainty in our extraction of the scaling exponent. See text for details.}
\label{ourpicture}
\end{figure}

In particular, we examined the temporal behavior of the spatial Wilson loop and demonstrated that it obeys an area law, with the scale set by a spatial string tension. We studied its temporal behavior in detail and extracted, for the first time, the power law that governs its decay with time. In analogy to the thermal case, the spatial string tension can be understood to determine the length scale for magnetic screening in the plasma. 

We next explored the dynamics of Chern-Simons charge in the Glasma. By employing two different cooling techniques we demonstrated the existence of integer valued topological transitions and studied their temporal evolution. In analogy to the thermal case, we computed the auto-correlation of the Chern-Simons charge; in contrast to the former, the auto-correlation function in the Glasma is non-Markovian and even demonstrates oscillatory behavior with increasing auto-correlation time. 

We argued that one can still identify a meaningful sphaleron transition rate for short auto-correlation times and studied the behavior of this rate with the evolution of the Glasma. Our first observation was that, with increasing cooling time, the Glasma sphaleron rate converges to a limiting value. This indicates that one is measuring the rate of genuine topological transitions at the maximal cooling times studied. Secondly, we found that while the sphaleron transition rate is sensitive to the initial occupancy at early times, it is insensitive to it at late times where it approaches a characteristic scaling behavior. Most strikingly, we find that the sphaleron transition rate, to a good approximation, scales with time as the string tension squared, or as the fourth power of the inverse magnetic screening length. 

While in a weakly coupled plasma in thermal equilibrium the magnetic screening length $l_{mag}$ is parametrically $\sim 1/\alpha_S T$, it is much smaller in the Glasma where initially $\l_{mag}$ is on the order of the inverse hard scale $1/Q_s$. As a consequence, the sphaleron transition rate in the Glasma is greatly enhanced compared to the equilibrium value of $\Gamma_{\rm eq} \sim \alpha_S^5 T^4$; in particular there is no parametric suppression of the sphaleron rate in the Glasma as $\Gamma_{\rm neq} \sim Q_s^4$. 

As the Glasma evolves and begins to develop a separation of scales, the sphaleron transition rate decreases, as illustrated in Fig.~\ref{ourpicture}. Within our classical-statistical framework we can follow this evolution up to a time scale $t_{\rm Quantum}\sim \alpha_S^{-7/4}\;Q_s^{-1}$ where the occupancy of hard modes in the plasma becomes of order unity and the classical description breaks down~\cite{Berges:2013fga,}. Employing our extraction of the time evolution of the string tension, we can estimate that by the time $t_{\rm Quantum}$ the sphaleron rate has dropped to $\Gamma_{\rm neq}\sim \alpha_S^{10/3} T^4$. In arriving at this estimate, we have equated the energy density of the Glasma to an equilibrium plasma with the same energy density, which gives us $T^4 \sim Q_s^4/\alpha_S$. Thus at $t_{\rm Quantum}$,  $\Gamma_{\rm neq}$ is still parametrically $\alpha_S^{-5/3}$ larger than the equilibrium value. 

There is non-trivial quantum dynamics at the end of the classical regime which prevents our following the evolution of the sphaleron transition rate all the way to equilibrium. While there has been a lot of progress in understanding the equilibration of hard modes based on kinetic theory~\cite{Kurkela:2014tea}, it will be important to study the equilibration process of the sphaleron rate in this quantum regime. 

From the perspective of computing the chiral magnetic effect, it is however the early time dynamics that matters the most. This is because the sphaleron transition rate is much larger at early times than the rate in thermal equilibrium. While this is encouraging, more work is required to understand the extrapolation of the weak coupling rates to realistic couplings as well as the sensitivity of the sphaleron transition rate to details of the initial conditions. This includes extending our simulations to SU(3) and to the longitudinally expanding Glasma realized in heavy ion collisions. A further essential improvement to the framework introduced here will include the addition of  Wilson fermions to the Glasma dynamics \cite{Borsanyi:2008eu, Saffin:2011kc,Gelfand:2016prm}. This will allow us to study the anomaly in real time. Finally, adding an external $U(1)$ electro-magnetic field to the fermion computation will allow us to investigate the Chiral Magnetic Effect in heavy ion collisions from first principles. The computation discussed here is the first necessary step in achieving the goals of this program. 

\section*{Acknowledgments}
We would like to thank Peter Arnold, J\"{u}rgen Berges, Dima Kharzeev, Guy D. Moore, Alexander Rothkopf, and Sayantan Sharma for helpful discussions. SS and RV are supported under DOE Contract No. DE-SC0012704. This research used the resources of the National Energy Research Scientific Computing Center, which is supported by the Office of Science of the U.S. Department of Energy under Contract No. DEAC02-05CH11231. MM, SS, and RV would like to thank the Institut f\"{u}r Theoretische Physik at the Universit\"{a}t Heidelberg for hospitality during the latter portion of this work. RV would like to thank the Excellence Initiative of Heidelberg University for their support. SS gratefully acknowledges a Goldhaber Distinguished Fellowship from Brookhaven Science Associates.

\appendix
\section{Calibrated Cooling Method}
\label{app::cooling}
We shall describe here further details of our implementation of the ``calibrated cooling" method developed  in \cite{Ambjorn:1997jz,Moore:1998swa}  to measure the Chern-Simons number. We focus on the aspects relevant to the practical implementation and refer to Sec \ref{calibratedcoolingsection} for a more general introduction.\\

\textit{Cooling.} We first perform a cooling of the spatial gauge links $U_\mu(x,t)$ at a given real time $t$ by following the energy gradient flow.  We introduce a dimensionless cooling time variable $\tau a^2$, which measures the depth of cooling in lattice units and perform a sequence of update steps of the gauge links according to
\begin{equation}
\label{eq:coolingstep}
U_\mu(x,t;\tau+\delta\tau)=\exp\big(-i g a E^{cool}_\mu(x,t;\tau)\delta\tau \big)U_\mu(x,t,\tau)\;,
\end{equation}
where the ``cooling" electric field $E^{cool}_\mu$ associated with the change of the gauge links along the cooling path is given by
\begin{eqnarray}
&&E^{cool,a}_i(x,t;\tau) = \\ 
&& \qquad \qquad - \frac{2}{g a^3} \sum_{j\neq i} \text{ReTr}\big[i\tau^a(U_{i,j}^{\Box}-U_{i,-j}^{\Box})(x,t;\tau)\big]\;,\nonumber
\end{eqnarray}
with the elementary plaquettes defined as
\begin{eqnarray}
\label{eq:plaquettes-def}
U_{i,j}^{\Box}(x) &=& U_i(x)U_j(x+\hat{i})U^\dagger_i(x+\hat{j})U^\dagger_j(x)\;,  \\ 
U_{i,-j}^{\Box}(x) &=& U^\dagger_j(x-\hat{j}) U_i(x-\hat{j}) U_j(x+\hat{i}-\hat{j}) U^{\dagger}_i(x)\;, \nonumber\\
U_{-i,j}^{\Box}(x)&=&U_{j}(x)U_{i}^{\dagger}(x+\hat{j}-\hat{i})U_{j}^{\dagger}(x-\hat{i})U_{i}(x-\hat{i})\;,  \nonumber  \\
U_{-i,-j}^{\Box}(x)&=&U_{i}^{\dagger}(x-\hat{i})U_{j}^{\dagger}(x-\hat{i}-\hat{j})U_{i}(x-\hat{i}-\hat{j})U_{j}(x-\hat{j})\;.  \nonumber
\end{eqnarray}
We use a cooling step size of $\delta\tau = a^2/8$ and repeat the update in Eq.~(\ref{eq:coolingstep}) until reaching the desired cooling depth $\tau_{c}$. During the course of the real time evolution, the cooling process is repeated after each $t=a/2$. Based on the cooled copies $U_\mu(x,t_1;\tau_c)$ and $U_\mu(x,t_2;\tau_c)$ of the original gauge field configurations at two adjacent times $t_{1}$ and $t_{2}$ (see Fig.~(\ref{fig:CoolingCartoon})) we then compute the change in the Chern-Simons number as detailed below.\\

\begin{figure}[t!]
\centering
\includegraphics[width=0.45\textwidth,natwidth=610,natheight=642]{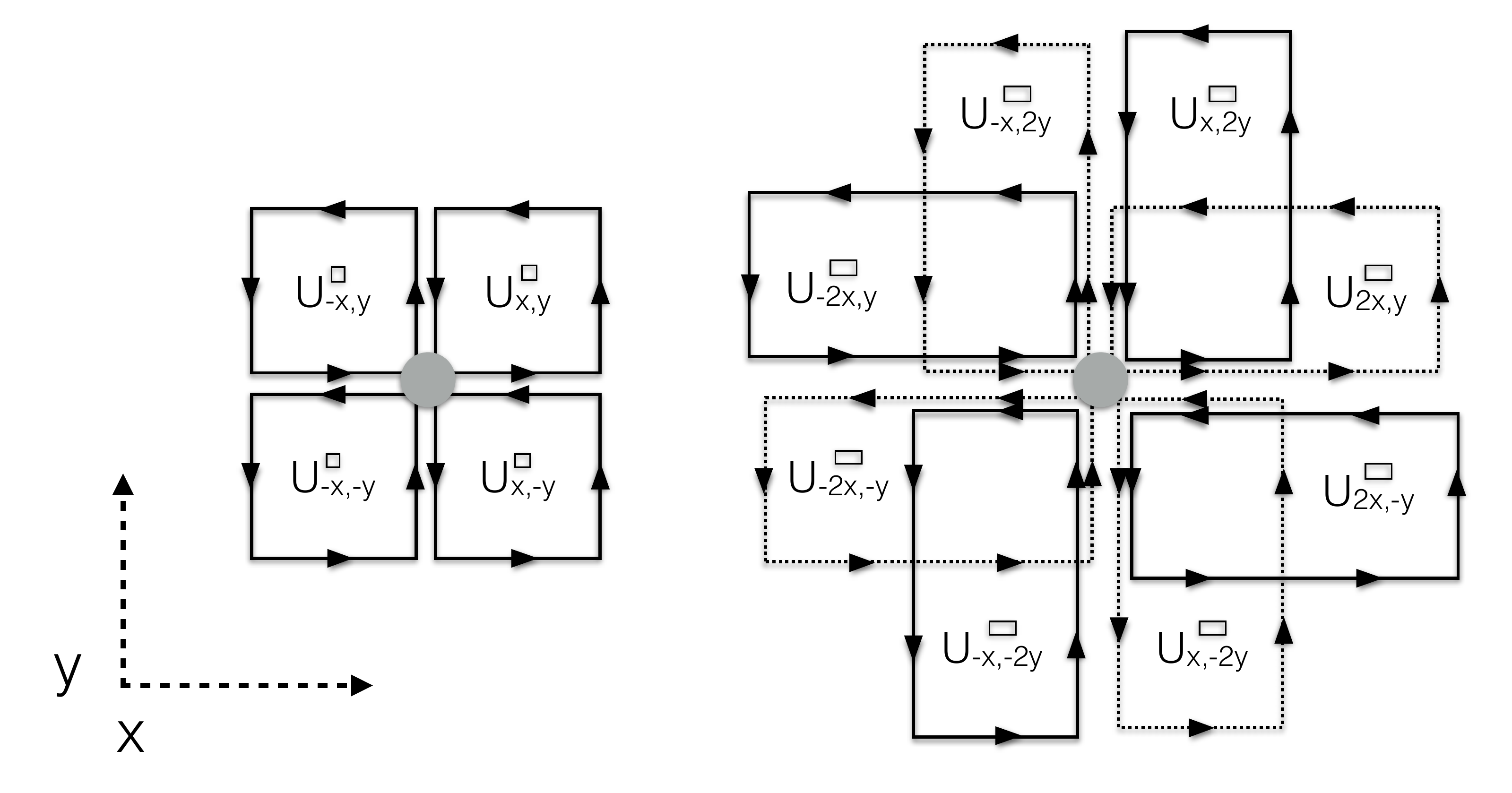}
\caption{Illustration of elementary square plaquettes (left) and rectangular plaquettes (right) employed in the computation of the magnetic field strength. }
\label{fig:LatticeOperators}
\end{figure}

\textit{Chern-Simons number.} Since we only perform cooling of the spatial gauge links, we first need to re-construct the connection between the cooled configurations $U_\mu(x,t_1;\tau_c)$ and $U_\mu(x,t_2;\tau_c)$ to evaluate the Chern-Simons current. Because the topology measurement does not rely on the exact path connecting the two configurations we do so by  simply performing a smooth interpolation between the starting point $U_\mu(x,t_1;\tau_c)$ and end point $U_\mu(x,t_2;\tau_c)$. Explicitly, we choose the cooled analogue of the color electric field
\begin{eqnarray}
E^{t_1 \rightarrow t_2}_i(x;\tau_c)=\frac{i}{ga (t_2-t_1)} \text{Log}\Big[U_i(x,t_2;\tau_c)U^\dagger_i(x,t_1;\tau_c)\Big] \nonumber \\
\end{eqnarray}
to be constant between the two adjacent times such that for $t_1\leq t \leq t_2$ the gauge links follow the trajectory
\begin{eqnarray}
\label{eq:coolinterpolation}
&&U_\mu(x,t_1\leq t \leq t_2;\tau_c)= \\
&& \qquad \exp\Big( -i g a  E^{t_1 \rightarrow t_2}_i(x;\tau_c)  (t-t_1) \Big) U_\mu(x,t_1;\tau_c)\;.  \nonumber
\end{eqnarray}
We then compute the change in Chern-Simons number between the two configurations by evaluating the space time integral of the Chern-Simons current
\begin{eqnarray}
\label{eq:DeltaNCsCCLattice}
&& N_{CS}^{\tau_c}(t_2)-N^{\tau_c}_{CS}(t_1)=\frac{g^2 a^3 (t_2-t_1)}{8 \pi^2} \sum_{x} E^{t_1 \rightarrow t_2}_{i,\rm{imp}}(x;\tau_c)  \nonumber  \\
 && \frac{B_{i}^{\rm{imp}}(x,t_1;\tau_c) +4 B_{i}^{\rm{imp}}(x,t_{\rm{mid}};\tau_c)+B_{i}^{\rm{imp}}(x,t_2;\tau_c)}{6} \nonumber  \\
\end{eqnarray}
where in order to improve the accuracy of the integral over time we construct the magnetic fields at the mid-point $t_{\rm{mid}}=(t_1+t_2)/2$ from Eq.~(\ref{eq:coolinterpolation}) and use  Simpson's rule to approximate the integral. Similarly, we use an $\mathcal{O}(a^2)$ improved definition of the color electric and color magnetic fields \cite{Moore:1996wn}, where the electric fields are locally determined at each lattice point according to
\begin{eqnarray}
&&E_{i,\rm{imp}}^{a}(x)= -\frac{1}{12}U_{i}^{ab}(x)E^{b}_i(x+\hat{i})  +\frac{7}{12} U_{i}^{\dagger,ab}(x-\hat{i}) E^{b}_i(x-\hat{i})  \nonumber  \\
&& +\frac{7}{12}E^a_i(x) -\frac{1}{12}U_{i}^{\dagger,ab}(x-\hat{i})U_{i}^{\dagger,bc}(x-2\hat{i})E_i^{c}(x-2\hat{i}) \, .
\end{eqnarray}
Here $U^{ab}=2~\text{tr}[t^{a}U t^{b} U^{\dagger}]$ denote the adjoint parallel transporters. Similarly, the magnetic fields are constructed from a combination of the four elementary (1x1) plaquettes in Eq.~(\ref{eq:plaquettes-def}) and the eight adjacent rectangular (2x1) plaquettes according to
\begin{eqnarray}
B_{i,\rm{imp}}^{a}(x)&=&\frac{\epsilon^{ijk}}{ga^2}~\text{ReTr}~\Big(i\tau^{a}\Big[\frac{5}{3}\sum_{4 \Box}U_{\pm j,\pm k}^{\Box}(x) \\
&& \qquad \qquad \qquad \qquad -\frac{1}{3}\sum_{8 \fbox{\rule{1.5pt}{-0.5pt}}} U_{\pm j,\pm k}^{\fbox{\rule{1.5pt}{-0.5pt}}}(x)\Big]\Big), \nonumber
\end{eqnarray} 
with the different lattice operators illustrated in Fig.~\ref{fig:LatticeOperators}.\\

\begin{figure}[t!]
\centering
\includegraphics[width=0.4\textwidth,natwidth=610,natheight=642]{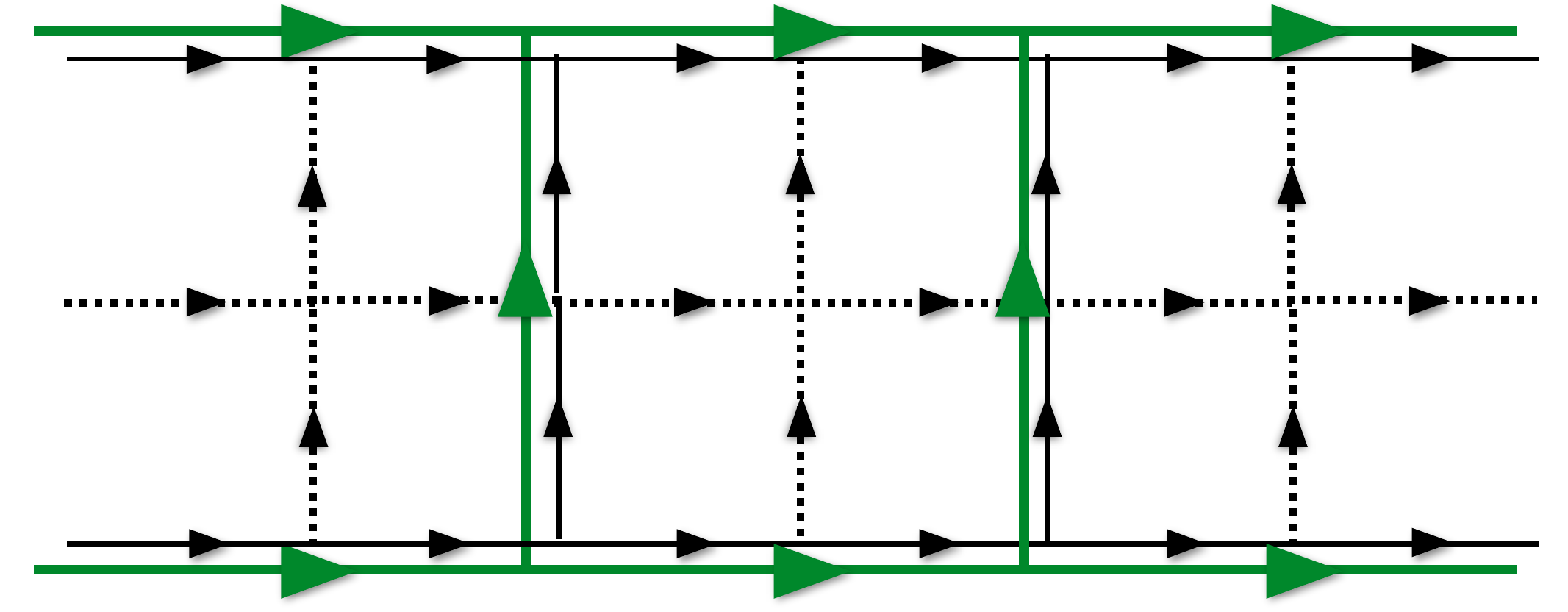}
\caption{Illustration of blocking procedure used only during the calibration step. A subset of the original gauge links represented by solid black lines is combined into a coarser lattice of green links.  }
\label{fig:Blocking}
\end{figure}

\textit{Calibration.} While the successive application of Eq.~(\ref{eq:DeltaNCsCCLattice}) allows us to follow the change of the Chern-Simons number over the course of the real time evolution,  it is useful in this process to re-calibrate the measurement occasionally to ensure that residual errors do not accumulate over time. As illustrated in Figure~\ref{fig:CoolingCartoon}, we perform additional calibration steps where we cool from $\tau_c$ all the way to the vacuum. Since the Chern-Simons number of the associated vacuum configuration is always an integer, we can get an independent estimate of the Chern-Simons number $N^{\tau_c}_{CS}(t)$ of the cooled configuration according to
\begin{eqnarray}
N^{\tau_c,\rm{calib}}_{CS}(t)=N_{CS}^{vac}(t)- \Delta N_{CS}^{\rm{coolingpath}}(t) \;,
\end{eqnarray}
which can used to re-calibrate the measurement. Here $\Delta N_{CS}^{\rm{cooling path}}(t)$ denotes the change of difference in Chern-Simons number computed along the cooling path from $\tau=\tau_c$ to the vacuum ($\tau\to \infty$)
\begin{eqnarray}
&&\Delta N_{CS}^{\rm{coolingpath}}(t)= \\
&& \qquad \qquad  \frac{g^2 a^3}{8 \pi^2} \int_{\tau_c}^{\infty} d\tau \sum_{x} E^{cool,a}_i(x,t;\tau) B^{cool,a}_i(x,t;\tau)\;. \nonumber
\end{eqnarray}
Since cooling all the way to the vacuum is computationally expensive, we follow earlier works and use blocking to reduce the numerical cost of the calibration procedure. Each time blocking is performed, neighboring sets of gauge links are combined into new ``blocked" links as illustrated in Fig~\ref{fig:Blocking}, which reduces the size of the lattice by a factor of $2^3$. Since  the effective step width $\delta \tau$ of the cooling can also be increased by a factor of $2^2$ after blocking, the numerical benefit is enormous and we have used up to two levels of blocking when performing calibration on our largest lattices. When sufficient cooling $\tau a^2 \gtrsim 1$ is performed before each level of blocking we find that the error introduced in the computation of $\Delta N_{CS}^{\rm{coolingpath}}(t)$  due to blocking can be kept below the 1\% percent level.\\ 

\begin{figure}[t!]
\centering
\includegraphics[width=0.45\textwidth,natwidth=610,natheight=642]{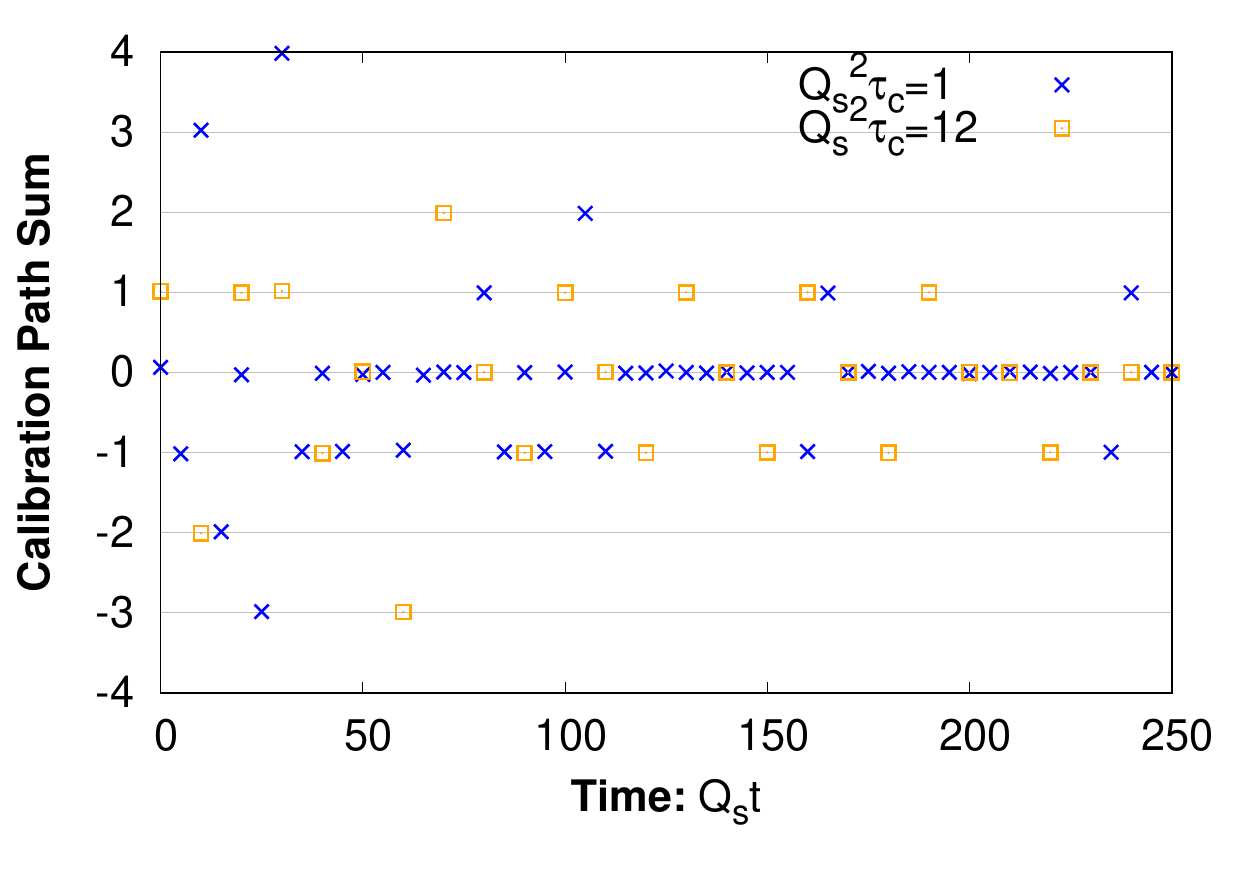}
\caption{Space-time integral of the Chern-Simons current computed along a path connecting two vacuum configurations according to the right hand side of Eq.~(\ref{eq:CoolingDerCheck}).  Data obtained for a single $N=96$, $Q_sa=1$ non-equilibrium configuration.}
\label{calibrationcheck}
\end{figure}

By combining the results $\Delta N_{CS}^{\rm{coolingpath}}(t_1)$ and $\Delta N_{CS}^{\rm{coolingpath}}(t_2)$ of consecutive calibrations with the measurement of the Chern-Simons number difference $N_{CS}^{\tau_c}(t_2)-N^{\tau_c}_{CS}(t_1)$ at the original cooling depth in Eq.~(\ref{eq:DeltaNCsCCLattice}), one can form a trajectory in configuration space which connects the vacuum configurations at $t_1$ and $t_2$ such that
\begin{eqnarray}
\label{eq:CoolingDerCheck}
&&N_{CS}^{vac}(t_2)-N_{CS}^{vac}(t_1)=-\Delta N_{CS}^{\rm{coolingpath}}(t_1)\\
&&  \qquad \qquad +N_{CS}^{\tau_c}(t_2)-N^{\tau_c}_{CS}(t_1)+\Delta N_{CS}^{\rm{coolingpath}}(t_2)\;. \nonumber
\end{eqnarray}
Because the difference in Chern-Simons number of two vacuum configurations $N_{CS}^{vac}(t_1)-N_{CS}^{vac}(t_2)$ on the left hand side is supposed to be an integer, this procedure allows for an explicit check of whether the lattice definition of the Chern-Simons current indeed behaves as a total derivative.  When evaluating the right hand side of Eq.~(\ref{eq:CoolingDerCheck}), we find that the deviation from integer values is typically less than 2 \%. An example of this calibration check is shown Figure~\ref{calibrationcheck}, where we plot the values obtained for the right hand side of Eq.~(\ref{eq:CoolingDerCheck}) over the coarse of the non-equilibrium evolution of a single configuration. Excellent agreement with integers can be observed, demonstrating that the measurement is indeed topological.

\section{Slave field}
\label{app::sf}
We will describe in this appendix the implementation of the slave field method \cite{Moore:1997cr}. As discussed in Sec.~\ref{slavefieldsection}, it provides an alternative measurement of the integer (topological) part of the Chern-Simons number. The basic philosophy underlying the slave field method is to identify the topological component of the Chern-Simons number with the winding number of the gauge transformation to the topologically trivial sector.  The challenge for this method however is to find this gauge transformation at every instance of time and ensure that it is sufficiently slowly varying to allow for a reliable extraction of the winding number.\\

\textit{Initialization \& Slave field dynamics}  When performing a slave field measurement, we perform an initial gauge fixing to set the dynamical gauge links and electric field variables such that they satisfy the (minimal) Coulomb gauge condition at initial time $t=0$. Since after the gauge fixing the configuration is topologically trivial, the initial condition for the slave field then simply becomes $S(x)=\id$.
Over the course of the real time evolution of the configuration the Coulomb gauge condition will be violated and one needs to update the slave field to dynamically keep track of the transformation back to Coulomb gauge. We follow the original reference \cite{Moore:1997cr} and use an update algorithm based on local lattice gauge fixing techniques to maximize the gauge fixing functional  $\sum_{x} \sum_i \text{ReTr}~w(x,t+\delta t)$ where
\begin{eqnarray}
&&w(x,t+\delta t)=S(x,t+\delta t) \sum_{i} \Big( U_{i}(x,t+\delta t)  S^{\dagger}(x+\hat{i},t+\delta t)  \nonumber \\
&& \qquad \qquad \qquad + U_{i}^{\dagger}(x-\hat{i} ,t+\delta t) S^{\dagger}(x-\hat{i},t+\delta t) \Big)\;,
\end{eqnarray}
Starting from an initial guess $S^{\rm{init}}(x,t+\delta t)$ for the slave field $S(x,t+\delta t)$, we perform $N_{steps}$ update steps of the Los Alamos gauge fixing algorithm, setting  \cite{Cucchieri:1995pn}
\begin{eqnarray}
S(x,t+\delta t) \rightarrow S^{\text{new}}(x,t+\delta t)= \tilde{w}^{\dagger}(x,t) S(x,t+\delta t)\;, \nonumber \\
\end{eqnarray}
in each step, where $w(x,t)$ is determined from the slave field in the previous step and $\tilde{w}$ denotes the projection of $w$ to $SU(N_c)$, which in the $SU(2)$ case simply takes the form $\tilde{w}=w/\sqrt{det(w)}$. 

In order to minimize the number of gauge fixing steps needed to reach acceptable gauge fixing precision, we follow \cite{Moore:1997cr} and try to take advantage of the previous update, by setting
\begin{eqnarray}
S^{\rm{init}}(x,t+\delta t)=(S(x,t)S^\dagger(x,t-\delta t))^{m}S(x,t)
\end{eqnarray}
with $m=1-\delta t/a$ as our initial guess for the slave field, except when the previous step was so large that $\text{ReTr} \big(\id-S(x,t)S^\dagger(x,t-\delta t) \big) > 2 (1-m)^2$ where we use $S^{\rm{init}}(x,t+\delta t)=S(x,t)$ instead. However when the peak-stress 
\begin{eqnarray}
&&PS(t)=\max_{x}~\Big( 3 N_c -\frac{1}{2}\text{ReTr}~w(x,t) \Big).
\end{eqnarray}
of the previous slave field configuration is above our tolerance $PS(t)>PS_{\rm{max}}$, we use $m=(1-\delta t/a)^3$ instead of the above and triple the number of gauge fixing steps $N_{steps}$. We found that for the small lattices used in our study of the thermal case, we achieve an accurate tracking of the gauge transformation with $N_{steps}=5$ and $PS_{\rm{max}}=1.2$. However for the larger lattices used in our non-equilibrium study, the local gauge fixing algorithm becomes inefficient and a much larger number of steps is needed. Unfortunately this makes the slave field method computationally too expensive to be of practical use when studying a large number of configurations on large lattices.

Even though the update described above is sufficient to determine the evolution of the slave field, it does not necessarily ensure that (over the course of the real time evolution) the slave field remains sufficiently slowly varying to reliably determine its winding number. However, as pointed out in  \cite{Moore:1997cr}, the smoothness of the slave field can be restored by performing the actual gauge transformation back to Coulomb gauge\begin{eqnarray}
U_{i}(x) \to U_{i}^{(S)}(x)\;, \quad  E_{i}(x) \to E_{i}^{(S)}(x)\;, \quad S(x) \to \id \;, \nonumber \\
\end{eqnarray}
when the peak stress is sufficiently small $PS(t)<PS_{\rm{max}}$. In practice, we check after every fifth time step whether this criterion is satisfied and eventually perform the transformation. We also note that, since performing the gauge transformation removes a possible winding, we have to add the winding number of the gauge transformation $S(x,t)$ to all subsequent measurements of the winding number. When the slave field is sufficiently slowly varying, it is then straightforward to determine its winding number using the methodology described in~\cite{Woit:1985jz,Moore:1997cr}.

\vspace{-0.5cm}

\bibliography{OESTR.bib}

\end{document}